\documentclass[preprint2]{proto}
\usepackage{times}
\usepackage{graphicx}
\newcommand{\refs}{\par\noindent\hangindent=1pc\hangafter=1}
\def\dominiketalsubsubsection#1{\vspace*{1ex}\noindent\refstepcounter{subsubsection}{\em
\thesubsubsection{} #1.}}

\def\um{$\mu$m}
\def\Df{\ensuremath{D_\mathrm{f}}}
\def\agr{a}
\def\rhog{\rho_\mathrm{g}}
\def\rhos{\rho_\mathrm{s}}
\def\rhop{\rho_\mathrm{p}}
\def\tstop{t_\mathrm{s}}
\def\cs{c_\mathrm{s}}
\def\fcontact{F_\mathrm{c}}
\def\vkep{V_\textrm{K}}

\begin{document}

\title{\textbf{\LARGE Growth of Dust as the Initial Step Toward Planet Formation}}

\author{\textbf{\large Carsten Dominik}}
\affil{\small\em Universiteit van Amsterdam}
\author{\textbf{\large J\"urgen Blum}}
\affil{\small\em Technische Universit\"at Braunschweig}
\author{\textbf{\large Jeffrey N. Cuzzi}}
\affil{\small\em NASA Ames Research Center}
\author{\textbf{\large Gerhard Wurm}}
\affil{\small\em Universit\"at M\"unster}

\begin{abstract}
\baselineskip = 11pt
\leftskip = 0.65in
\rightskip = 0.65in
\parindent=1pc
{\small We discuss the results of laboratory
  measurements and theoretical models concerning the aggregation of
  dust in protoplanetary disks, as the initial step toward planet
  formation.  Small particles easily stick when they collide and form
  aggregates with an open, often fractal structure, depending on the
  growth process.  Larger particles are still expected to grow at
  collision velocities of about 1m/s.  Experiments also show that,
  after an intermezzo of destructive velocities, high collision
  velocities above 10m/s on porous materials again lead to net growth
  of the target.  Considerations of dust-gas interactions show that
  collision velocities for particles not too different in
  surface-to-mass ratio remain limited up to sizes about 1m, and
  growth seems to be guaranteed to reach these sizes quickly and
  easily.  For meter sizes, coupling to nebula turbulence makes
  destructive processes more likely.  Global aggregation models show
  that in a turbulent nebula, small particles are swept up too fast to be
  consistent with observations of disks.  An extended phase may
  therefore exist in the nebula during which the small particle
  component is kept alive through collisions driven by turbulence
  which frustrates growth to planetesimals until conditions are more
  favorable for one or more reasons.
 \\~\\~\\~}
 
\end{abstract}

\section{\textbf{INTRODUCTION}}

The growth of dust particles by aggregation stands at the beginning of
planet formation.  Whether planetesimals form by incremental
aggregation, or through gravitational instabilities in a dusty
sublayer, particles have to grow and settle to the midplane regardless.
On the most basic level, the physics of such growth is simple: Particles
collide because relative velocities are induced by random and
(size-dependent) systematic motions of grains and aggregates in the
gaseous nebula surrounding a forming star.  The details are, however,
highly complex.  The physical state of the disk, in particular the
presence or absence of turbulent motions, set the boundary conditions.
When particles collide with low velocities, they stick by mutual
attractive forces, be it simple van der Waals attraction or stronger
forces (molecular dipole interaction in polar ices, or grain-scale
long-range forces due to charges or magnetic fields).  While the lowest
velocities create particle shapes governed by the motions alone,
larger velocities contribute to shaping the aggregates by
restructuring and destruction.  The ability to internally dissipate
energy is critical in the growth through intermediate pebble and
boulder sizes.  In this review we will concentrate on the physical
properties and growth characteristics of these small and intermediate
sizes, but also make some comments on the formation of planetesimals.

Relative velocities between grains in a protoplanetary disk can be
caused by a variety of processes.  For the smallest grains, these are
dominated by Brownian motions, that provide relative velocities in the
mm/s to cm/s range for (sub)micron sized grains.  Larger grains show
systematic velocities in the nebula because they decouple from the
gas, settle vertically, and drift radially.  At 1AU in a solar nebula,
these settling velocities reach m/s for cm-sized grains.  Radial drift
becomes important for even larger particles and reaches 10's of m/s
for m-sized bodies.  Finally, turbulent gas motions can induce
relative motions between particles.  For details see for example
\textit{Weidenschilling} (1977; 1984), \textit{Weidenschilling and
  Cuzzi} (1993), \textit{Cuzzi and Hogan} (2003).

The timescales of growth processes and the density and strength of
aggregates formed by them, will depend on the structure of the
aggregates.  A factor of overriding importance for dust--gas
interactions (and therefore for the timescales and physics of
aggregation), for the stability of aggregates, and for optical properties
alike is the structure of aggregates as they form through the
different processes.

The interaction of particles with the nebula gas is determined
primarily by their gas drag stopping time $\tstop$ which is given by
\begin{equation}
\label{eq:3}
\tstop = \frac{mv}{F_\mathrm{fric}}
=\frac{3}{4\cs\rhog} \frac{m}{\sigma}
\end{equation}
where $m$ is the mass of a particle, $v$ its velocity relative to the
gas, $\sigma$ the average projected surface area, $\rhog$ is the
gas density, $\cs$ is the sound speed, and $F_\mathrm{fric}$ is the
drag force.  The second equal sign in eq.(\ref{eq:3}) holds under the
assumption that particles move at sub-sonic velocities and that the
mean free path of a gas molecule is large compared to the size of the
particle (Epstein regime).  In this case, the stopping time is
proportional to the ratio of mass and cross section of the particle.
For spherical non-fractal (i.e.  compact or porous) particles of
radius $\agr$ and mass density $\rhos$, this can be written as $\tstop =
\agr \rhos / c \rhog$.  Fractal particles are characterized by the
fact that the average density of a particle depends on size in a
powerlaw fashion, with a power (the fractal dimension $\Df$) smaller
than 3.
\begin{equation}\label{mass-size}
    m(\agr) \propto \agr^{\Df} ~.
\end{equation}
For large aggregates, this value can in principle be measured for
individual particles.  For small particles, it is often more
convenient to measure it using sizes and masses of a distribution of
particles.  

\begin{figure}[t!]
\epsscale{1.0}
\plotone{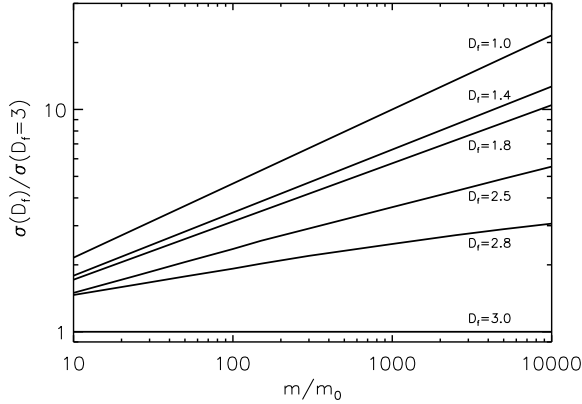}
\caption{\label{fig:sigma}\small Projected area of aggregates as a function
  of aggregate size and fractal dimension, normalized to the cross
  section of a compact particle with the same mass.}
\end{figure}

Fractal particles generally have large surface-to-mass ratios; in the
limiting case of long linear chains ($\Df=1$) of grains with radii
$\agr_0$, $\sigma/m$ approaches the constant value
$3\pi/(16\agr_0\rhos)$.  This value differs from the value for a
single grain $3/(4\agr_0\rhos)$ by just a factor $\pi/4$.
Fig.~\ref{fig:sigma} shows how the cross section of particles varies
with their mass for different fractal dimensions.  It shows that for
aggregates made of 10000 monomers, the surface-to-mass ratio can
easily differ by a factor of 10.  An aggregate made from 0.1\um{}
particles with a mass equivalent to a 10\um{} particle consists of
10$^6$ monomers and the stopping time could vary by a factor of order
100.  Just how far the fractal growth of aggregates proceeds is really
not yet known.

This review is organized as follows: In
section~\ref{sec:dust-aggr-exper} we cover the experiments and theory
describing the basic growth processes of dust aggregates.  In
section~\ref{sec:part-gas-inter} we discuss particle-gas interactions
and the implications for inter-particle collision velocities as well as
planetesimal formation.  In section~\ref{sec:global-disk-models} we
describe recent advances in the modeling of dust aggregation in
protoplanetary disks and observable consequences.

\section{\textbf{DUST AGGREGATION EXPERIMENTS AND THEORY}}
\label{sec:dust-aggr-exper}
\subsection{Interactions between individual dust grains}

\dominiketalsubsubsection{Interparticle adhesion forces}
\label{adhesion force}
Let us assume that the dust grains are spherical in shape and that
they are electrically neutral and non-magnetic. In that case, two
grains with radii $\agr_1$ and $\agr_2$ will always experience a
short-range attraction due to induced dielectric forces, e.g. van der
Waals interaction. This attractive force results in an elastic
deformation leading to a flattening of the grains in the contact
region. An equilibrium is reached when the attractive force equals
the elastic repulsion force. For small, hard grains with low
surface forces, the equilibrium contact force is given by
(\textit{Derjaguin et al.}, 1975)
\begin{equation}\label{dmt}
    \fcontact = 4 \pi \gamma_\mathrm{s} R ~,
\end{equation}
where $\gamma_\mathrm{s}$ and $R$ denote the specific surface energy of the grain
material and the local radius of surface curvature, given by $R = \agr_1
\agr_2 / (\agr_1 + \agr_2)$, respectively.  Measurements of the separation
force between pairs of $\rm SiO_2$ spheres with radii $\agr$ between
0.5\,\um{} and 2.5\,\um{} (corresponding to reduced radii $R = 0.35
\ldots 1.3$\,\um{}) confirm the validity of Eq. \ref{dmt} (\textit{Heim et
al.}, 1999).

\dominiketalsubsubsection{\label{rolling force}Interparticle rolling-friction forces}
Possibly the most important parameter influencing the structure of
aggregates resulting from low velocity collisions is the
resistance to rolling motion.  If this resistance is very strong,
both aggregate compaction and internal energy dissipation in
aggregates would be very difficult.  Resistance to rolling first
of all depends strongly on the geometry of the grains.  If grains
contain extended flat surfaces, contact made on such locations
could not be moved by rolling - any attempt to roll them would
inevitably lead to breaking the contact.  In the contact between
round surfaces, resistance to rolling must come from an asymmetric
distribution of the stresses in the contact area. Without external
forces, the net torque exerted on the grains should be zero.
\textit{Dominik and Tielens} (1995) showed that the pressure
distribution becomes asymmetric, when the contact area is slightly
shifted with respect to the axis connecting the curvature centers
of the surfaces in contact.  The resulting torque is
\begin{equation}
\label{eq:1}
M = 4\fcontact \left(\frac{a_\mathrm{contact}}{a_{\mathrm{contact,0}}}\right)^{3/2} \xi
\end{equation}
where $a_\mathrm{contact,0}$ is the equilibrium contact radius, $a_\mathrm{contact}$ the actual contact
radius due to pressure in the vertical direction, and $\xi$ is the
displacement of the contact area due to the torque.  In this picture,
energy dissipation, and therefore friction, occurs when the contact
area suddenly readjusts after it has been displaced because of
external forces acting on the grains.  The friction force is
proportional to the pull-off force $\fcontact$.

\textit{Heim et al.} (1999) observed the reaction of a chain of dust
grains using a long-distance microscope and measured the applied
force with an Atomic Force Microscope (AFM). The derived
rolling-friction forces between two $\rm SiO_2$ spheres with radii
of $\agr= 0.95~\rm \mu m$ are $F_\mathrm{roll} = (8.5 \pm 1.6) \cdot
10^{-10}$ N.  If we recall that there are two
grains involved in rolling, we get for the rolling-friction
energy, defined through a displacement of an angle $\pi/2$
\begin{equation}\label{eroll}
    E_\mathrm{roll} = \pi \agr F_\mathrm{roll} = {\rm O}(10^{-15} {\rm J}) ~.
\end{equation}
Recently, the rolling of particle chains has been observed under
the scanning electron microscope while the contact forces were
measured simultaneously (\textit{Heim et al.}, 2005).

\dominiketalsubsubsection{\label{sticking efficiency}Sticking efficiency in single grain collisions}
The dynamical interaction between small dust grains was derived by
\textit{Poppe et al.} (2000a) in an experiment in which single,
micrometer-sized dust grains impacted smooth targets at various
velocities ($0 \ldots 100$\,m/s) under
vacuum conditions.
For spherical grains,
a sharp transition from sticking with an efficiency of $\beta
\approx 1$ to bouncing (i.e. a sticking efficiency of $\beta =
0$) was observed.  This threshold velocity is $v_\mathrm{s} \approx 1.2~ \rm
m/s$ for $\agr = 0.6$\,\um{} and $v_\mathrm{s} \approx 1.9~ \rm
m/s$ for $\agr = 0.25$\,\um.
It decreases with increasing grain size. The target materials were
either polished quartz or atomically-smooth (surface-oxidized)
silicon.  Currently, no theoretical explanation is available
for the threshold velocity for sticking. Earlier attempts to model
the low-velocity impact behavior of spherical grains predicted
much lower sticking velocities (\textit{Chokshi et al.}, 1993).  These
models are based upon impact experiments with ``softer''
polystyrene grains (\textit{Dahneke}, 1975). The main difference becomes
visible when studying the behavior of the rebound grains in
non-sticking collisions.
In the experiments by \textit{Dahneke} (1975) and also in those by
\textit{Bridges et al.} (1996) using macroscopic ice grains, the behavior of
grains after a bouncing collision was a unique function of the impact
velocity, with a coefficient of restitution (rebound velocity divided
by impact velocity) always close to unity and increasing monotonically
above the threshold velocity for sticking.
For harder, still spherical, $\rm SiO_2$ grains (\textit{Poppe et al.},
2000a), the \emph{average} coefficient of restitution decreases
considerably with increasing impact velocity.  In addition to
that, \emph{individual} grain impacts show considerable scatter in
the coefficient of restitution.

The impact behavior of irregular dust grains is more complex.
Irregular grains of various sizes and compositions show an overall
decrease in the sticking probability with increasing impact velocity.
The transition from $\beta = 1$ to $\beta = 0$, however, is very broad
so that even impacts as fast as $v \approx 100~ \rm m/s$ can lead
to sticking with a moderate probability.

\subsection{Dust aggregation and restructuring}

\dominiketalsubsubsection{\label{laboratory aggregation}Laboratory and microgravity aggregation
experiments}
In recent years, a number of laboratory and microgravity
experiments have been carried out to derive the aggregation
behavior of dust under conditions of young planetary systems. To
be able to compare the experimental results to theoretical
predictions and to allow numerical modelling of growth phases that
are not accessible to experimental investigation, ``ideal''
systems were studied, in which the dust grains were monodisperse
(i.e. all of the same size) and initially non-aggregated.  
Whenever the mean collision velocity between the dust grains or
aggregates is much smaller than the sticking threshold (see section
\ref{sticking efficiency}), the aggregates formed in the experiments
are ``fractal'', i.e. $\Df<3$ (\textit{Wurm and Blum}, 1998;
\textit{Blum et al.}, 1999; \textit{Blum et al.}, 2000; \textit{Krause
  and Blum}, 2004).  The precise value of the fractal dimension
depends on the specific aggregation process and can reach values as
low as $\Df = 1.4$ for Brownian-motion driven aggregation
(\textit{Blum et al.}, 2000; \textit{Krause and Blum}, 2004;
\textit{Paszun and Dominik}, 2006), $\Df = 1.9$ for aggregation in a
turbulent gas (\textit{Wurm and Blum}, 1998), or $\Df = 1.8$ for
aggregation by gravitationally driven sedimentation in gas
(\textit{Blum et al.}, 1999). It is inherent to a dust aggregation
process in which aggregates with low fractal dimensions are formed
that the mass distribution function is rather narrow
(quasi-monodisperse) at any given time. In all realistic cases, the
mean aggregate mass $\bar{m}$ follows either a power law with time
$t$, i.e. $\bar{m} \propto t^\gamma$ with $\gamma > 0$ (\textit{Krause
  and Blum}, 2004) or grows exponentially fast, $\bar{m} \propto
\exp(\delta t)$ with $\delta >0$ (\textit{Wurm and Blum}, 1998) which
can be verified in dust-aggregation models (see Section
\ref{modelling}).
\par
\begin{figure}[t!]
\epsscale{1.0} \plotone{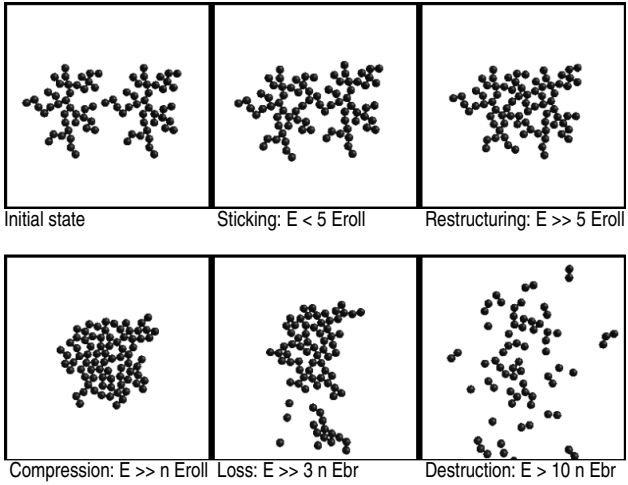}
\caption{\label{fig:restruct}\small Dominating processes and
  associated energies in aggregate collisions, after \textit{Dominik
    and Tielens} (1997) and \textit{Wurm and Blum} (2000).
  $E_\mathrm{br}$ is the energy needed to break a contact,
  $E_\mathrm{roll}$ is the energy to roll two grains through an angle
  $\pi/2$ and $n$ is the total number of contacts in the aggregates.}
\end{figure}

As predicted by \textit{Dominik and Tielens} (1995, 1996, 1997), experiments
have shown that at collision velocities near the velocity threshold
for sticking (of the individual dust grains), a new phenomenon occurs
(\textit{Blum and Wurm}, 2000). Whereas at low impact speeds, the
aggregates' structures are preserved in collisions (the so-called
``hit-and-stick'' behavior), the forming aggregates are compacted at
higher velocities.  In even more energetic collisions, the aggregates
fragment so that no net growth is observable. The different stages of
compaction and fragmentation are depicted in Fig.~\ref{fig:restruct}.

\dominiketalsubsubsection{Modelling of dust aggregation}
\label{modelling}
The evolution of grain morphologies and masses for a system of
initially monodisperse spherical grains that are subjected to Brownian
motion has been studied numerically by \textit{Kempf et al.} (1999).
The mean aggregate mass increases with time following a power law (see
Section  \ref{laboratory aggregation}).  The aggregates have fractal
structures with a mean fractal dimension of $\Df = 1.8$. Analogous
experiments by \textit{Blum et al.} (2000) and \textit{Krause and
  Blum} (2004), however, found that the mean fractal dimension was
$\Df = 1.4$. Recent numerical work by \textit{Paszun and Dominik} (2006) showed
that this lower value is caused by Brownian rotation (neglected by
\textit{Kempf et al.} (1999)).  More chain-like dust aggregates can form if the
mean free path of the colliding aggregates becomes smaller than their
size, i.e. if the assumption of ballistic collisions breaks down and a
random-walk must be considered for the approach of the particles.  The
fractal dimension of thermally aggregating dust grains is therefore
dependent on gas pressure and reaches an asymptotic value of $\Df =
1.5$ for the low density conditions prevailing through most of the
presolar nebula.  Only in the innermost regions the densities are high
enough to cause deviations.

The experimental work reviewed in Section  \ref{laboratory aggregation}
can be used to test the applicability of
theoretical dust aggregation models.  Most commonly, the mean-field
approach by \textit{Smoluchowski} (1916) is used for the description of the
number density $n(m,t)$ of dust aggregates with mass $m$ as a function
of time $t$.  Smoluchowski's rate equation reads in the integral form
\begin{eqnarray}\label{smoluchowski1}
    \frac{\partial n(m,t)}{\partial t} & = & \frac{1}{2} \int_0^m
    K(m',m-m')\\
    && \cdot n(m',t)~n(m-m',t)~{\rm d}m' \nonumber\\
    &&- n(m,t)\int_0^\infty
    K(m',m)~n(m',t)~{\rm d}m'~. \nonumber
\end{eqnarray}
Here, $K(m_1,m_2)$ is the reaction kernel for aggregation of the
coagulation equation \ref{smoluchowski1}. The first term on the
rhs. of Eq. \ref{smoluchowski1} describes the rate of sticking
collisions between dust particles of masses $m'$ and $m-m'$ whose
combined masses are $m$ (gain in number density for the mass $m$).
The second term
denotes a loss in the number density for the mass $m$ due to
sticking collisions between particles of mass $m$ and mass $m'$.
The factor 1/2 in the first term
accounts for the fact that each pair collision is counted twice in the
integral. In most astrophysical applications the gas densities are so
low that dust aggregates collide ballistically. In that case, the
kernel in Eq.  \ref{smoluchowski1} is given by
\begin{equation}
        K(m_1,m_2) = \beta(m_1,m_2;v) ~ v(m_1,m_2) ~ \sigma(m_1,m_2) ~,
        \label{kernel1}
\end{equation}
where $\beta(m_1,m_2;v)$, $v(m_1,m_2)$, and $\sigma(m_1,m_2)$ are
the sticking probability, the collision velocity, and the cross
section for collisions between aggregates of masses $m_1$ and
$m_2$, respectively.

A comparison between numerical predictions from Eq.
\ref{smoluchowski1} and experimental results on dust aggregation
was given by \textit{Wurm and Blum} (1998) who investigated dust
aggregation in rarefied, turbulent gas. Good agreement for both
the mass distribution functions and the temporal behavior of the
mean mass was found when using a sticking probability of
$\beta(m_1,m_2;v)=1$, a mass-independent relative velocity between
the dust aggregates and the expression by \textit{Ossenkopf} (1993) for the
collision cross section of fractal dust aggregates.  \textit{Blum} (2006)
showed that the mass distribution of the fractal aggregates
observed by \textit{Krause and Blum} (2004) for Brownian-motion driven
aggregation can also be modelled in the transition regime between
free-molecular and hydrodynamic gas flow.

Analogous to the experimental findings for the collisional behavior of
fractal dust aggregates with increasing impact energy (\textit{Blum
  and Wurm} (2000), see Section \ref{laboratory aggregation}),
\textit{Dominik and Tielens} (1997) showed in numerical experiments on
aggregate collisions that with increasing collision velocity the
following phases can be distinguished: hit-and-stick behavior,
compaction, loss of monomer grains, and complete fragmentation (see
Fig.~\ref{fig:restruct}).
They also showed that the outcome of a collision depends on the impact
energy, the rolling-friction energy (see Eq. \ref{eroll} in Section
\ref{rolling force}) and the energy for the breakup of single
interparticle contacts (see Section \ref{adhesion force}).  The model by
\textit{Dominik and Tielens} (1997) was quantitatively confirmed by
the experiments of \textit{Blum and Wurm} (2000) (see Fig.~\ref{fig:restruct}).

To analyze observations of protoplanetary disks and model the
radiative transfer therein, the optical properties of particles are
important (\textit{McCabe et al.}, 2003; \textit{Ueta and Meixner},
2003; \textit{Wolf}, 2003). Especially for particle sizes comparable
to the wavelength of the radiation, the shape and morphology of a
particle are of major influence for the way the particle interacts
with the radiation. With respect to this, it is important to know how
dust evolution changes the morphology of a particle. As seen above, in
most cases dust particles are not individual monolithic solids but
rather aggregates of primary dust grains. Numerous measurements and
calculations have been carried out on aggregates (e.g. \textit{Kozasa
  et al.}, 1992; \textit{Henning and Stognienko}, 1996; \textit{Wurm
  and Schnaiter}, 2002; \textit{Gustafson and Kolokolova}, 1999;
\textit{Wurm et al.}, 2004a; \textit{Min et al.}, 2005).  No simple
view can be given within the frame of this paper.  However, it is
clear that the morphology and size of the aggregates will strongly
influence the optical properties.

\dominiketalsubsubsection{Aggregation with long-range forces}
Long range forces may play a role in the aggregation process, if
grains are either electrically charged or magnetic.  Small iron grains
may become spontaneously magnetic if they are single domain
(\textit{Nuth et al.}, 1994; \textit{Nuth and Wilkinson}, 1995),
typically at sizes of a few tens of nanometers.  Larger grains
containing ferromagnetic components can be magnetized by an impulse
magnetic field generated during a lightning discharge (\textit{T\'unyi
  et al.}, 2003).  For such magnetized grains, the collisional cross
section is strongly enhanced compared to the geometrical cross section
(\textit{Dominik and N\"ubold}, 2002). Aggregates formed from magnetic
grains remain strong magnetic dipoles, if the growth process keeps the
grain dipoles aligned in the aggregate (\textit{N\"ubold and
  Glassmeier}, 2000).  Laboratory experiments show the spontaneous
formation of elongated, almost linear aggregates, in particular in the
presence of an external magnetic field (\textit{N\"ubold et al.},
2003).  The relevance of magnetic grains to the formation of
macroscopic dust aggregates is, however, unclear.

Electric charges can be introduced through tribo-electric effects in
collisions, through which electrons and/or ions are exchanged between
the particles (\textit{Poppe et al.}, 2000b; \textit{Poppe and
  Schr\"apler}, 2005; \textit{Desch and Cuzzi}, 2000). The number of
separated elementary charges in a collision between a dust particle
and a solid target with impact energy $E_\mathrm{c}$ can be expressed by
(\textit{Poppe et al.}, 2000b; \textit{Poppe and Schr\"apler}, 2005)
\begin{equation}\label{charging}
    N_e \approx \left(\frac{E_\mathrm{c}}{10^{-15} {\rm J}}\right)^{0.8} ~.
\end{equation}
The cumulative effect of many non-sticking collisions can lead to an
accumulation of charges and to the build-up of strong electrical
fields at the surface of a larger aggregate. In this way, impact
charging could lead to electrostatic trapping of the impinging dust
grains or aggregates (\textit{Blum}, 2004). Moreover, impact charging
and successive charge separation can cause an electric discharge in
the nebula gas.  For nebula lighting (\textit{Desch and Cuzzi}, 2000) a
few hundred to thousand elementary charges per dust grain are
required.  This corresponds to impact velocities in the range $20
\ldots 100 ~\rm m/s$ (\textit{Poppe and Schr\"apler}, 2005) which
seems rather high for mm particles.

Electrostatic attraction by dipole-dipole forces has been seen to be
important for grains of several hundred micron radius (chondrule size)
forming clumps that are centimeters to tens of centimeters across
(\textit{Marshall et al.}, 2005; \textit{Ivlev et al.}, 2002).  Spot
charges distributed over the grain surfaces lead to a net dipole of
the grains, with growth dynamics very similar to that of magnetic
grains.  Experiments in microgravity have shown spontaneous
aggregation of particles in the several hundred micron size regime
(\textit{Marshall and Cuzzi}, 2001; \textit{Marshall et al.}, 2005;
\textit{Love and Pettit}, 2004).  The aggregates show greatly enhanced
stability, consistent with cohesive forces increased by factors of
10$^3$ compared to the normal van der Waals interaction.  Based on the
experiments,
for weakly charged dust grains, the electrostatic interaction energy
at contact for the charge-dipole interaction is in most cases larger
than that for the charge-charge interaction. For heavily-charged
particles, the mean mass of the system does not grow faster than
linearly with time, i.e. even slower than in the non-charged case for
Brownian motion (\textit{Ivlev et al.}, 2002; \textit{Konopka et al.},
2005).

\subsection{Growth and compaction of large dust aggregates}

\dominiketalsubsubsection{\label{physical properties of aggregates}Physical
  properties of macroscopic dust aggregates}
Macroscopic dust aggregates can be created in the laboratory by a
process termed random ballistic deposition (RBD, \textit{Blum and
  Schr\"apler}, 2004). In its idealized form, RBD uses individual,
spherical and monodisperse grains which are deposited randomly but
unidirectionally on a semi-infinite target aggregate.  The volume
filling factor $\phi=0.11$ of these aggregates, defined as the
fraction of the volume filled by dust grains, is identical to
ballistic particle-cluster aggregation which occurs when a bimodal
size distribution of particles (aggregates of one size and individual
dust grains) is present and when the aggregation rates between the
large aggregates and the small particles exceed those between all
other combinations of particle sizes. When using idealized
experimental parameters, i.e.  monodisperse spherical $\rm SiO_2$
grains with $0.75 ~ \rm \mu m$ radius, \textit{Blum and Schr\"apler}
(2004) measured a mean volume filling factor for their macroscopic
(cm-sized) RBD dust aggregates of $\phi = 0.15$, in full agreement
with numerical predictions (\textit{Watson et al.}  1997). Relaxing
the idealized grain morphology resulted in a decrease of the volume
filling factor to values of $\phi = 0.10$ for quasi-monodisperse,
irregular diamond grains and $\phi = 0.07$ for polydisperse, irregular
$\rm SiO_2$ grains (\textit{Blum}, 2004).

Static uniaxial compression experiments with the macroscopic RBD dust
aggregates consisting of monodisperse spherical grains (\textit{Blum
  and Schr\"apler}, 2004) showed that the volume filling factor remains
constant as long as the stress on the sample is below $\sim 500 ~ \rm
N~m^{-2}$. For higher stresses, the volume filling factor
monotonically increases from $\phi = 0.15$ to $\phi = 0.34$. Above
$\sim 10^5 ~ \rm N~m^{-2}$, the volume filling factor remains constant
at $\phi = 0.33$. Thus, the compressive strength of the uncompressed
sample is $\Sigma \approx 500 ~ \rm N~m^{-2}$.  These values differ
from those derived with the models of \textit{Greenberg et al.} (1995)
and \textit{Sirono and Greenberg} (2000) by a factor of a few. The compressive
strengths of the macroscopic dust aggregates consisting of irregular
and polydisperse grains was slightly lower at $\Sigma \sim 200 ~ \rm
N~m^{-2}$. The maximum compression of these bodies was reached
for stresses above $\sim 5 \cdot 10^5 ~ \rm N~m^{-2}$ and resulted in volume
filling factors as low as $\phi = 0.20$ (\textit{Blum}, 2004). As a maximum
compressive stress of $\sim 10^5 \ldots 10^6~ \rm N~m^{-2}$
corresponds to impact velocities of $\sim 15 \ldots 50 ~\rm m/s$
which are typical for meter-sized protoplanetary dust aggregates, we
expect a maximum volume filling factor for these bodies in the solar
nebula of $\phi = 0.20 \ldots 0.34$. \textit{Blum and Schr\"apler} (2004) also
measured the tensile strength of their aggregates and found for
slightly compressed samples ($\phi = 0.23$) $T = 1,000 ~ \rm
N~m^{-2}$. Depending on the grain shape and the size distribution, the
tensile strength decreased to values of $T \sim 200 ~ \rm
N~m^{-2}$ for the uncompressed case (\textit{Blum}, 2004).

\textit{Sirono} (2004) used the above continuum properties of macroscopic
dust aggregates, i.e. compressive strength and tensile strength,
to model the collisions between protoplanetary dust aggregates.
For sticking to occur in an aggregate-aggregate collision, \textit{Sirono}
(2004) found that the impact velocity must follow the relation
\begin{equation}\label{sound speed}
    v < 0.04 \sqrt{\frac{{\rm d \Sigma(\phi)}}{{\rm d
    \rho(\phi)}}}~,
\end{equation}
where $\rho(\phi) = \rho_0 \cdot \phi$ is the mass density of the
aggregate and $\rho_0$ denotes the mass density of the grain
material. Moreover, the conditions $\Sigma(\phi) < Y(\phi)$ and
$\Sigma(\phi) < T(\phi)$ must be fulfilled. For the shear
strength, \textit{Sirono} (2004) applies $Y(\phi) = \sqrt{2 \Sigma(\phi)
T(\phi) / 3}$. A low compressive strength of the colliding
aggregates favors compaction and, thus, damage restoration which
can otherwise lead to a break-up of the aggregates. In addition, a
large tensile strength also prevents the aggregates from being
disrupted in the collision.

\textit{Blum and Schr\"apler} (2004) found an approximate relation
between compressive strength and volume filling factor
\begin{equation}\label{relcompvff}
    \Sigma(\phi) = \Sigma_\mathrm{s} \left( \phi - \phi_0 \right)^{0.8} ~,
\end{equation}
which is valid in the range $\phi_0 = 0.15 \le \phi \le 0.21$.  Such a
scaling law was also found for other types of macroscopic aggregates,
e.g. for jammed toner particles in fluidized bed experiments
(\textit{Valverde et al.}, 2004). For the aggregates consisting of
monodisperse $\rm SiO_2$ spheres, the scaling factor $\Sigma_\mathrm{s}$ can be
determined to be $\Sigma_\mathrm{s} = 2.9 \cdot 10^4 ~ \rm N~m^{-2}$. If we
apply Eq. \ref{relcompvff} to Eq. \ref{sound speed}, we get, with
$\rho(\phi) = \rho_0 \cdot \phi$ and $\rho_0 = 2 \cdot 10^3 ~ {\rm
  kg~m^{-3}}$, for the impact velocity of low-density dust aggregates
\begin{equation}\label{sound speed 2}
    v < 0.04 \sqrt{\frac{0.8 \Sigma_\mathrm{s}}{\rho_0 (\phi -
    \phi_0)^{0.2}}} \approx 0.14  ~ (\phi -\phi_0)^{-0.1} ~ {\rm m/s}~.
\end{equation}
Although the function in Eq. \ref{sound speed 2} goes to infinity
for $\phi \rightarrow \phi_0$, for all practical purposes the
characteristic velocity is strongly restricted. For volume filling
factors $\phi \ge 0.16$ we get $v < 0.22 ~ {\rm m/s}$. Thus,
following the SPH simulations by \textit{Sirono} (2004), we expect
aggregate sticking in collisions for impact velocities $v \lesssim
0.2 ~ {\rm m/s}$.

\begin{figure}[t!]
\epsscale{1.0} \plotone{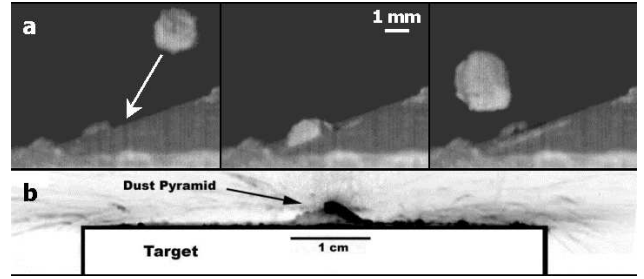}
\caption{\label{fig:impact}\small (a) Non-sticking (v=1.8\,m/s)
  oblique impact between high-porosity dust aggregates
  (\textit{Langkowski and Blum}, unpublished data). The three images
  show, from left to right, the approaching projectile aggregate
  before, during, and after impact.  The arrow in the left image
  denotes the impact direction of the projectile aggregate. The
  experiment was performed under microgravity conditions. It is
  clearly visible that the rebounding aggregate (right image) is more
  massive than before the collision.  (b) Result of a high-velocity
  normal impact ($v=23.7$\,m/s) between compacted aggregates (\textit{Wurm et
  al.}, 2005). About half of the projectile mass sticks to the target
  after the impact and is visible by its pyramidal structure on the
  flat target. Mind the different size scales in (a) and (b).}
\end{figure}

\dominiketalsubsubsection{\label{low_velocity_collisions}
Low-velocity collisions between macroscopic dust aggregates} Let us
now consider recent results in the field of high-porosity aggregate
collisions. \textit{Langkowski and Blum} (unpublished data) performed
microgravity collision experiments between 0.1-1 mm-sized (projectile)
RBD aggregates and 2.5 cm-sized (target) RBD aggregates. Both
aggregates consisted of monodisperse spherical $\rm SiO_2$ grains with
radii of $\agr = 0.75 ~ \rm \mu m$. In addition to that, impact
experiments with high-porosity aggregates consisting of irregular
and/or polydisperse grains were performed. The parameter space of the
impact experiments by Langkowski and Blum encompassed collision
velocities in the range $0 < v < 3 ~ {\rm m/s}$ and projectile
masses of $10^{-9} ~{\rm kg} \le m \le 5 \cdot 10^{-6} ~ {\rm kg}$ for
all possible impact parameters (i.e. from normal to tangential
impact).
Surprisingly, through most of the parameter space, the collisions did
lead to sticking.  The experiments with aggregates consisting of
monodisperse spherical $\rm SiO_2$ grains show, however, a steep
decrease in sticking probability from $\beta=1$ to $\beta=0$ if the
tangential component of the impact energy exceeds $\sim 10^{-6}$ J
(see the example of a non-sticking impact in Fig.~\ref{fig:impact}a).
Other materials also show the tendency towards lower sticking
probabilities with increasing tangential impact energies. As these
aggregates are ``softer'', the decline in sticking probability in the
investigated parameter space is not complete. When the projectile
aggregates did not stick to the target aggregate, considerable mass
transfer from the target to the projectile aggregate takes place
during the impact (\textit{Langkowski and Blum}, unpublished data).
Typically, the mass of the projectile aggregate was doubled after a
non-sticking collision (see Fig.~\ref{fig:impact}a).

The occurrence of sticking in aggregate-aggregate collisions at
velocities $\gtrsim 1~\rm m/s$ is clearly in disagreement
with the prediction by \textit{Sirono} (2004) (see Eq. \ref{sound speed}).
In addition, the evaluation of the experimental data shows that
the condition for sticking, $\Sigma(\phi) < Y(\phi)$, seems not to
be fulfilled for high-porosity dust aggregates. This means that
the continuum aggregate model by \textit{Sirono} (2004) is still not
precise enough to fully describe the collision and sticking
behavior of macroscopic dust aggregates.

\dominiketalsubsubsection{\label{high_velocity_collisions}High-velocity
collisions between macroscopic dust aggregates}
The experiments described above indicate that at velocities above
approximately 1\,m/s, collisions turns from sticking to bouncing, at
least for oblique impacts.  At higher velocities one would naively
expect that bouncing and eventually erosion will continue to dominate,
and this is also observed in a number of different experiments
(\textit{Colwell}, 2003; \textit{Bridges et al.}  1996, \textit{Kouchi
  et al.}, 2002; \textit{Blum and M\"unch}, 1993; \textit{Blum and
  Wurm}, 2000).

Growth models which assume sticking at velocities $\gg 1 ~\rm
m/s$ are therefore often considered to be impossible (e.g.
\textit{Youdin}, 2004).  As velocities $\ga 10\rm ~m/s$ clearly
occur for particles that have exceeded m-size, this is a fundamental
problem for the formation of planetesimals.

However, recent experiments (\textit{Wurm et al.}, 2005) have studied
impacts of mm-sized compact dust aggregates onto cm-sized compact
aggregate targets at impact velocities between 6 and 25 $\rm m/s$.
Compact aggregates can be the result of previous sticking or
non-sticking collisions (see Sections \ref{physical properties of
  aggregates} and \ref{low_velocity_collisions}).  Both projectile and
target consisted of $\rm \mu m$-sized dust particles. In agreement
with the usual findings at lower impact velocities around a few $\rm
m/s$, the projectiles just rebound, slightly fracture or even remove
some parts of the target.  However, as the velocity increases
{\it{above}} a threshold of 13 $\rm m/s$, about half of the mass of
the projectile rigidly sticks to the target after the collision while
essentially no mass is removed from the target (see
Fig.~\ref{fig:impact}b). Obviously, higher collision velocities can be
favorable for growth, probably by destroying the internal structure of
the porous material and dissipating energy in this way.  \par Only
about half of the impactor contributes to the growth of the target in
the experiments.  The other half is ejected in the form of small
fragments, with the important implication that these collisions both
lead to net growth of the target and return small particles to the
disk. This keeps dust abundant in the disk over a long time. For the
specific experiments by \textit{Wurm et al.} (2005), the fragments
were evenly distributed in size up to 0.5 mm.  In a certain sense, the
disk might thus quickly turn into a ``debris disk'' already at early
times.  We will get back to this point in section
\ref{sec:glob-settl-aggr}.

\section{PARTICLE-GAS INTERACTION}
\label{sec:part-gas-inter}
Above we have seen that small solid particles grow rapidly into
aggregates of quite substantial sizes, while retaining their fractal
nature (in the early growth stage) or a moderate to high porosity (for
later growth stages).  From the properties of primitive meteorites, we
have a somewhat different picture of nebula particulates - most of the
solids (chondrules, CAIs, metal grains, etc) were individually compacted
as the result of unknown melting processes, and were highly
size-sorted. Even the porosity of what seem to be fine-grained
accretion rims on chondrules is 25\% or less (\textit{Scott et al.},
1996; \textit{Cuzzi}, 2004; \textit{Wasson et al.}, 2005). Because
age-dating of chondrules and chondrites implies a delay of a Myr or
more after formation of the first solids, it seems possible that, in
the asteroid formation region at least, widespread accretion to parent
body sizes did not occur until after the mystery melting events began
which formed the chondrules.

It may be that conditions differed between the inner and outer solar
system.  Chondrule formation might not have occurred at all in the
outer solar system where comet nuclei formed, so some evidence of the
fractal aggregate growth stage may remain in the granular structure of
comet nuclei. New results from Deep Impact imply that comet Tempel 1
has a porosity of 60-80\% (\textit{A'Hearn et al.}, 2005)!  This value
is in agreement with similar porosities found in several other comets
(\textit{Davidsson}, 2006).  Even in the terrestrial planet/asteroid
belt region, there is little reason to doubt that growth of aggregates
started well before the chondrule formation era, and continued into
and (probably) throughout it.  Perhaps, after chondrules formed,
previously ineffective growth processes might have dominated (sections
3.2 and 3.3).

\subsection{Radial and vertical evolution of solids}
\dominiketalsubsubsection{Evolution prior to formation of a dense midplane layer} 
\label{sec:prior-to-midplane}
The nebula gas (but not the particles) experiences radial pressure
gradients because of changing gas density and temperature. These
pressure gradients act as small modifications to the central gravity
from the star that dominates orbital motion, so that the gas and
particles orbit at different speeds and a gas drag force exists
between them which constantly changes their orbital energy and angular
momentum. Because the overall nebula pressure gradient force is
outward, it counteracts a small amount of the inward gravitational
force and the gas generally orbits more slowly than the particles, so
the particles experience a headwind which saps their orbital energy,
and the dominant particle drift is inward. Early work on gas-drag
related drift was by \textit{Whipple} (1972), \textit{Adachi et al.}
(1976), and \textit{Weidenschilling} (1977). Analytical solutions for
how particles interact with a non-turbulent nebula having a typically
outward pressure gradient were developed by \textit{Nakagawa et al.}
(1986).  For instance, the ratio of the pressure gradient force to the
dominant central gravity is $\eta \sim 2 \times 10^{-3}$, leading to a
net velocity difference between the gas and particles orbiting at
Keplerian velocity $\vkep$ of $\eta \vkep$ (see, e.g. \textit{Nakagawa et
  al.}  1986). However, if local radial maxima in gas pressure exist,
particles will drift towards their centers from both sides, possibly
leading to radial bands of enhancement of solids (see section 3.4.1).

Small particles generally drift slowly inwards, at perhaps a few
cm/s; even this slow inexorable drift has generated some concern
over the years, as to how CAIs (early, high-temperature condensates)
can survive over the apparent 1-3 Myr period between their creation
and the time they were incorporated into chondrite meteorite parent
bodies.  This concern, however, neglected the role of turbulent
diffusion (see Section ~\ref{sec:role-of-turb}). Particles of meter
size drift inwards very rapidly - 1 AU/century. It has often been
assumed that these particles were ``lost into the sun'', but more
realistically, their inward drift first brings them into regions warm
enough to evaporate their primary constituents, which then become
entrained in the more slowly evolving gas and increase in relative
abundance as inward migration of solids supplies material faster than
it can be removed. Early models describing significant global
redistribution of solids relative to the nebula gas by radial drift
were presented by \textit{Morfill and V{\"o}lk} (1984) and
\textit{Stepinski and Valageas} (1996, 1997); these models either
ignored midplane settling or made simplifying approximations regarding
it, and did not emphasize the potential for enhancing material in the
vapor phase. Indeed, however, because of the large mass fluxes
involved, this ``evaporation front'' effect can alter the nebula
composition and chemistry significantly (\textit{Cuzzi et al.}, 2003;
\textit{Cuzzi and Zahnle}, 2004; \textit{Yurimoto and Kuramoto}, 2004;
\textit{Krot et al.}, 2005; \textit{Ciesla and Cuzzi}, 2005); see also
\textit{Cyr et al.} (1999) for a discussion; however, the results of
this paper are inconsistent with similar work by Supulver and
\textit{Lin} (2000) and \textit{Ciesla and Cuzzi} (2005). This stage
can occur very early in nebula history, long before formation of
objects large enough to be meteorite parent bodies.

\dominiketalsubsubsection{The role of turbulence}
\label{sec:role-of-turb}
The presence or absence of gas turbulence plays a critical role in the
evolution of nebula solids. There is currently no widespread agreement
on just how the nebula gas may be maintained in a turbulent state
across all regions of interest, if indeed it is (\textit{Stone et
  al.}, 2000, \textit{Cuzzi and Weidenschilling}, 2005). Therefore we
discuss both turbulent and non-turbulent situations. For simplicity we
will treat turbulent diffusivity $\cal D$ as equal to turbulent
viscosity $\nu_\mathrm{T} = \alpha c H$, where $c$ and $H$ are the nebula sound
speed and vertical scale height, and $\alpha \ll 1$ is a
non-dimensional scaling parameter. Evolutionary timescales of observed
protoplanetary nebulae suggest that $10^{-5} < \alpha < 10^{-2}$ in
some global sense. The largest eddies in turbulence have scale sizes
$H \sqrt{\alpha}$ and velocities $v_\mathrm{turb} = c \sqrt{\alpha}$
(\textit{Shakura et al.}, 1978; \textit{Cuzzi et al.}, 2001).

Particles respond to forcing by eddies of different frequency and
velocity as described by \textit{V\"{o}lk et al.} (1980) and
\textit{Markiewicz et al.} (1991), determining their relative
velocities with respect to the gas and to each other. The diffusive
properties of MRI turbulence, at least, seem not to differ in any
significant way from the standard homogeneous, isotropic models in
this regard (\textit{Johansen and Klahr}, 2005). Analytical solutions
for resulting particle velocities in these regimes were derived by
\textit{Cuzzi and Hogan} (2003).  These are discussed in more detail
below and by \textit{Cuzzi and Weidenschilling} (2005).

Vertical turbulent diffusion at intensity $\alpha$ maintains particles
of stopping time $\tstop$ in a layer of thickness $h \sim H \sqrt{\alpha
  /\Omega \tstop}$ (\textit{Dubrulle et al.}, 1995; \textit{Cuzzi et
  al.}, 1996), or a solid density enhancement $H/h = \sqrt{\Omega \tstop
  / \alpha}$ above the average value. For particles of 10 cm size and
smaller and $\alpha > \alpha_\mathrm{min}=10^{-6}(\agr/1{\rm cm})$
(\textit{Cuzzi and Weidenschilling}, 2005), the resulting layer is
much too large and dilute for collective particle effects to dominate
gas motions, so radial drift and diffusion continue unabated.  Outward
radial diffusion relieves the long-standing worry about ``loss into
the sun" of small particles, such as CAIs, which are too small to
sediment into any sort of midplane layer unless turbulence is
vanishingly small ($\alpha \ll \alpha_\mathrm{min}$), and allows some fraction
of them to survive over 1 to several Myrs after their formation as
indicated by meteoritic observations (\textit{Cuzzi et al.}, 2003). A
similar effect might help explain the presence of crystalline
silicates in comets (\textit{Bockel\'ee-Morvan}, 2002; \textit{Gail},
2004).

\dominiketalsubsubsection{Dense midplane layers} 
\label{sec:dense-midplane-physics}
When particles {\it are} able to settle to the midplane, the particle density
gets large enough to dominate the motions of the local gas. This is the regime
of {\it collective effects}; that is, the behavior of a particle depends
indirectly on how all other local particles combined  affect the gas in which
they move. In regions where collective effects are important, the mass-dominant
particles can drive the entrained gas to orbit at nearly Keplerian
velocities (if they are sufficiently well coupled to the gas), and thus the
headwind the gas can exert upon the particles diminishes from $\eta \vkep$ 
(section 3.1.1). This causes the headwind-driven radial drift and all other
differential particle velocities caused by gas drag to diminish as well.

The particle mass loading $\rhop/\rhog$ cannot increase without
limit as particles settle, even if the global turbulence vanishes, and
the density of settled particle layers is somewhat self-limiting. The
relative velocity solutions of \textit{Nakagawa et al.} (1986) apply
in particle-laden regimes once $\rhop/\rhog$ is known, but do not
provide for a fully self consistent determination of $\rhop/\rhog$
in the above sense; this was addressed by \textit{Weidenschilling}
(1980) and subsequently \textit{Cuzzi et al.} (1993), \textit{Champney
  et al.} (1995), and \textit{Dobrovolskis et al.} (1999). The latter
numerical models are similar in spirit to the simple analytical
solutions of \textit{Dubrulle et al.} (1995) mentioned earlier, but
treat large particle, high mass loading regimes in globally
nonturbulent nebulae which the analytic solutions cannot address.
Basically, as the midplane particle density increases, local,
entrained gas is accelerated to near-Keplerian velocities by drag
forces from the particles. Well above the dense midplane, the gas
still orbits at its pressure-supported, sub-Keplerian rate. Thus there
is a vertical shear gradient in the orbital velocity of the gas, and
the velocity shear creates turbulence which stirs the particles.  This
is sometimes called ``self-generated turbulence". Ultimately a
steady-state condition arises where the particle layer thickness
reaches an equilibrium between downward settling and upward diffusion.
This effect acts to block a number of gravitational instability
mechanisms in the midplane (section ~\ref{sec:instab}).

\subsection{Relative velocities and growth in turbulent and nonturbulent
nebulae}
\label{sec:relat-veloc-growth}

In both turbulent and nonturbulent regimes, particle relative
velocities drive growth to larger sizes. Below we show that relative
velocities in both turbulent and nonturbulent regimes are probably
small enough for accretion and growth to be commonplace and rapid, at
least until particles reach meter size or so. We only present results
here for particles up to a meter or so in size, because the expression
for gas drag takes on a different form at larger sizes.  As particles
grow, their mass per unit area increases so they are less easily
influenced by the gas, and ``decouple'' from it. Their overall drift
velocities and relative velocities all diminish roughly in a linear
fashion with particle radius larger than a meter or so (\textit{Cuzzi
  and Weidenschilling}, 2005).

We use particle velocities relative to the gas as derived by
\textit{Nakagawa et al.} (1986) for a range of local particle mass
density relative to the gas density (their equations 2.11, 2.12, and
2.21) to derive particle velocities relative to each other in the same
environment; all relative velocities scale with $\eta \vkep$.

\begin{figure}[t!]
\epsscale{1.5}
\includegraphics[width=8.2cm]{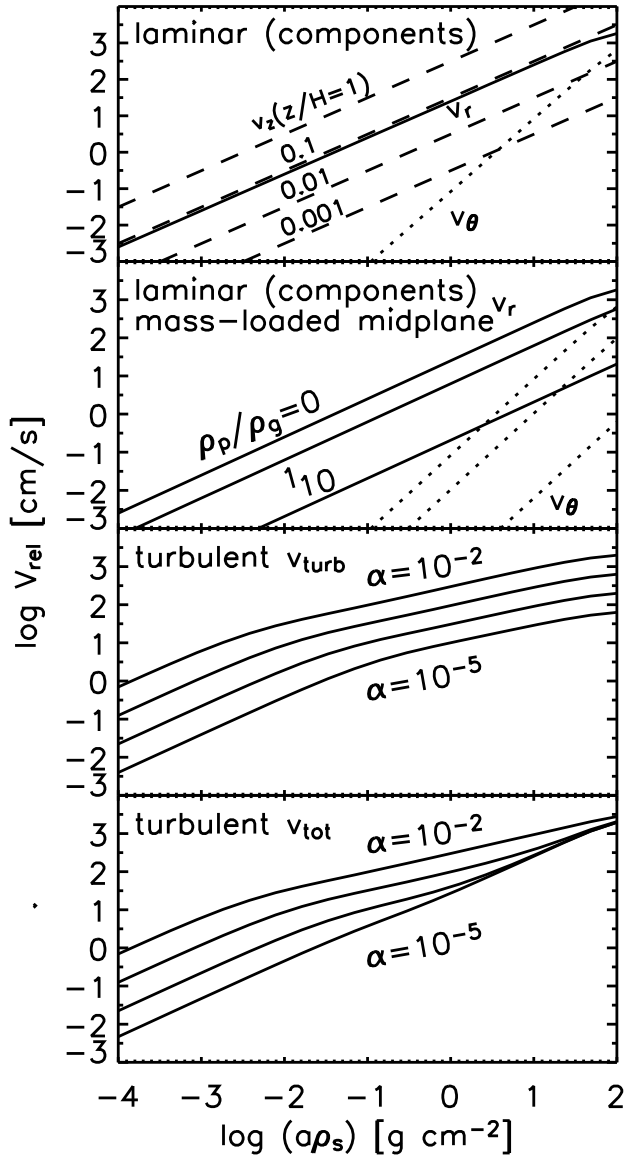}
\caption{\label{fig:vrel}\small Relative velocities between particles   of
  radii $\agr$ and $\agr/3$, in nebulae which are non-turbulent (top two
  panels) or turbulent (bottom two panels), for a minimum mass solar
  nebula at 2.5AU.  In the top panel, the particle density $\rhop$ is
  assumed to be much smaller than the gas density $\rhog$:
  $\rhop/\rhog \ll 1$. Shown are the radial (solid line) and
  azimuthal (dotted line) components of the relative velocities.  The
  dashed curves show the vertical relative velocities, which depend on
  height above the midplane and are shown for different values of
  $z/H$.  For nonturbulent cases, particles settle into dense midplane
  layers (section ~\ref{sec:dense-midplane-physics}), so a more
  realistic situation would be $\rhop/\rhog \ga 1$ or even $\gg 1$
  (\textit{Cuzzi et al.}, 1993); thus in the second panel we show relative
  radial and angular velocities for three different values of
  $\rhop/\rhog =$ 0, 1, and 10.  For these high mass loadings, $z/H$
  must be small, so the vertical velocities are smaller than the
  radial velocities. In the third panel we show relative velocities
  for the same particle size difference due only to turbulence, for
  several values of $\alpha$. In the bottom panel, we show the
  quadrature sum of turbulent and non-turbulent velocities, assuming
  $z/H=0.01$.}
\end{figure}

For simplicity we will assume particles which differ by a factor of three in
radius; \textit{Weidenschilling} (1997) finds mass accretion to be dominated by size
spreads on this order; the results are insensitive to this factor. Relative
velocities for particles of radii $\agr$ and $\agr/3$, in the absence of turbulence
and due only to differential, pressure-gradient-driven gas drag, are plotted in
the top two panels of Fig.~\ref{fig:vrel}.  In the top panel we show cases
where collective effects are negligible (particle density $\rhop$ $\ll$ gas
density $\rhog$). Differential vertical settling (shown at different heights
$z$ above the midplane, as normalized by the gas vertical scale height $H$)
dominates relative velocities and particle growth high above the midplane $(z/H
> 0.1)$, and radial relative velocities dominate at lower elevations. Except
for the largest particles, relative velocities for particles with this size
difference are much less than $\eta \vkep$; particles closer in mass would have
even smaller relative velocities.

Moreover, in a dense midplane layer, when collective effects dominate
(section~\ref{sec:dense-midplane-physics}), all these relative velocities are
reduced considerably from the values shown. In the second panel we show radial
and azimuthal relative velocities for several values of $\rhop/\rhog$. When
the particle density exceeds the gas density, collective effects reduce the
headwind, and all relative velocities diminish.

Relative velocities in turbulence of several different intensities, as
constrained by the nebula $\alpha$ (again for particles of radii $\agr$
and $\agr/3$), are shown in the two bottom panels. In the second panel
from the bottom, relative velocities are calculated as the difference
of their velocities relative to the turbulent gas, neglecting
systematic drifts and using analytical solutions derived by
\textit{Cuzzi and Hogan} (2003; their equation 20) to the formalism of
\textit{V\"{o}lk et al.} (1980). Here, the relative velocities are
forced by turbulent eddies with a range of size scales, having eddy
turnover times ranging from the orbit period (for the large eddies) to
much smaller values (for the smaller eddies), and scale with $v_\mathrm{turb}
= c \alpha^{1/2}$.

In the bottom panel we sum the various relative velocities in quadrature to get
an idea of total relative velocities in a turbulent nebula in which particles
are also evolving by systematic gas-pressure-gradient driven drift. This
primarily increases the relative velocities of the larger particles in the
lower $\alpha$ cases.

Overall, keeping in mind the critical velocities for sticking
discussed in section 2 ($\sim$ m/s), and that particle surfaces are
surely crushy and dissipative, one sees that for particles up to a
meter or so, growth by sticking is plausible even in turbulent nebulae
for a wide range of $\alpha$.  Crushy aggregates will grow by
accumulating smaller crushy aggregates as described in earlier
sections ({\it eg.} \textit{Weidenschilling}, 1997, for the laminar
case). After this burst of initial growth to roughly meter size,
however, the evolution of solids is very sensitive to the presence or
absence of global nebula turbulence, as described in sections 3.3-3.4
below. Meter-size particles inevitably couple to the largest eddies,
with $v_\mathrm{turb} \ga$ several meters per second, and would destroy each
other if they were to collide. We refer to this as the fragmentation
limit. However, if particles can somehow grow their way past 10 meters
in size, their survival becomes more assured because all relative
velocities, such as shown in Fig.~\ref{fig:vrel}, decrease linearly
with $\agr \rhos$ for values larger than shown in the plot due to the
linear decrease of the area/mass ratio.

\dominiketalsubsubsection{Another role of gas in growth beyond the
  fragmentation limit}
The role of gas in protoplanetary disks is not restricted to generate relative
velocities between two bodies which then collide. The gas also plays an
important role {\it during} individual collisions. A large body which moves
through the disk faces a headwind and collisions with smaller aggregates take
place at its front (headwind) side. Fragments are thus ejected against the wind
and can be driven back to the surface by the gas flow.

For small bodies the gas flow can be regarded as free molecular flow.
Thus streamlines end on the target surface and the gas drag is always
towards the surface.  Whether a fragment returns to the surface
depends on its gas-particle coupling time (i.e. size and density) and
on the ejection speed and angle.  Whether reaccretion of enough
fragments for net growth occurs, eventually depends on the
distribution of ejecta parameters, gas density, and target size.  It
was shown by \textit{Wurm et al.} (2001) that growth of a larger body
due to impact of dust aggregates entrained in a head wind is possible
for collision velocities above 12m/s. At 1 AU a 30-cm body in a disk
model according to \textit{Weidenschilling and Cuzzi} (1993) can grow in a
collision with small dust aggregates even if the initial collision is
rather destructive.

\textit{Sekiya and Takeda} (2003) and \textit{K\"unzli and Benz}
(2003) showed that the mechanism of aerodynamic reaccretion might be
restricted to a maximum size due to a change in the flow regime from
molecular to hydrodynamic. Fragments are then transported around the
target rather then back to it. \textit{Wurm et al.} (2004b) argue that
very porous targets would allow some flow going through the body,
which would still allow aerodynamic reaccretion, but this strongly
depends on the morphology of the body (\textit{Sekiya and Takeda},
2005). As the gas density decreases outwards in protoplanetary disks,
the maximum size for aerodynamic reaccretion increases. However, the
minimum size also increases and the mechanism is only important for
objects which have already grown beyond the fragmentation limit in
some other way - {\it e.g.} by immediate sticking of parts of larger
particles as discussed above (\textit{Wurm et al.}, 2005).

\subsection{Planetesimal formation in a midplane layer} 
\dominiketalsubsubsection{Incremental growth}
Based on relative velocity arguments such as given above,
\textit{Weidenschilling} (1988, 1997) and \textit{Dullemond and Dominik} (2004,
2005) find that growth to meter size is rapid (100-1000 yr at 1 AU;
6-7$\times 10^4$ yrs at 30 AU) whether the nebula is turbulent or not.
Such large particles settle towards the midplane within an orbit
period or so.  However, in turbulence, even meter-sized particles are
dispersed sufficiently that the midplane density remains low, and
growth remains slow. A combination of rapid radial drift, generally
erosive, high-velocity impacts with smaller particles, and occasional
destructive collisions with other meter-sized particles frustrates
growth beyond meter-size or so under these conditions.

In {\it nonturbulent} nebulae, even smaller particles can settle into
fairly thin midplane layers and the total particle densities can
easily become large enough for collective effects to drive the
entrained midplane gas to Keplerian, diminishing both headwind-induced
radial drift and relative velocities. In this situation, meter-sized
particles quickly grow their way out of their troublesome tendency to
drift radially (\textit{Cuzzi et al.}, 1993); planetesimal-sized
objects form in only $10^3 - 10^4$ years at 1 AU
(\textit{Weidenschilling}, 2000), and a few times $10^5$ years at 30 AU
(\textit{Weidenschilling}, 1997). However, such robust growth may, in
fact, be too rapid to match observations of several kinds (see
section~\ref{sec:glob-settl-aggr} and chapters by \textit{Dullemond et
  al.} and \textit{Natta et al.})

\dominiketalsubsubsection{Particle layer instabilities}
\label{sec:instab}
While to some workers the simplicity of ``incremental growth'' by
sticking in the dense midplane layer of a nonturbulent nebula is
appealing, past uncertainty in sticking properties has led others to
pursue instability mechanisms for particle growth which are
insensitive to these uncertainties. Nearly all instability mechanisms
discussed to date (\textit{Safronov}, 1969, 1991; \textit{Goldreich
  and Ward}, 1973; \textit{Ward}, 1976, 2000; \textit{Sekiya}, 1983,
1998; \textit{Goodman and Pindor}, 2000; \textit{Youdin and Shu}, 2002)
occur {\it only} in nebulae where turbulence is essentially absent,
and particle relative velocities are already very low. Just how low
the global turbulence must be depends on the particle size involved,
and the nebula $\alpha$ (sections~\ref{sec:prior-to-midplane} and
~\ref{sec:dense-midplane-physics}).

Classical treatments (the best known is \textit{Goldreich and Ward},
1973) assume that gas pressure plays no role in gravitational
instability, being replaced by an effective pressure due to particle
random velocities (below we note this is not the case). Particle
random velocities act to puff up a layer and reduce its density below
the critical value, which is always on the order of the so-called
Roche density $\rho^* \sim M_{\odot}/R^3$ where $R$ is the distance to
the central star; different workers give constraints which differ by
factors of order unity ({\it cf.} \textit{Goldreich and Ward}, 1973,
\textit{Weidenschilling}, 1980; \textit{Safronov}, 1991; \textit{Cuzzi
  et al.}, 1993). These criteria can be traced back through Goldreich
and \textit{Ward} (1973) to \textit{Goldreich and Lynden-Bell} (1965),
\textit{Toomre} (1964), \textit{Chandrasekhar} (1961) and
\textit{Jeans} (1928), and in parallel through \textit{Safronov}
(1960), \textit{Bel and Schatzman} (1958), and \textit{Gurevitch and
  Lebedinsky} (1950).  Substituting typical values one derives a
formal, nominal requirement that the local particle mass density must
exceed about $10^{-7}$ g cm$^{-3}$ at 2 AU from a solar mass star even
for {\it marginal} gravitational instability - temporary gravitational
clumping of small amplitude - to occur. This is about $10^3$ times
larger than the gas density of typical minimum mass nebulae, requiring
enhancement of the solids by a factor of about $10^5$ for a typical
average solids-to-gas ratio. From section ~\ref{sec:role-of-turb} we
thus require the particle layer to have a thickness $h < 10^{-5} H$,
which in turn places constraints on the particle random velocities $h
\Omega$ and on the global value of $\alpha$.

Even assuming global turbulence to vanish, \textit{Weidenschilling}
(1980, 1984) noted that turbulence stirred by the very dense particle
layer itself will puff it up to thicknesses $h$ that precluded even
this marginal gravitational instability.  This is because turbulent
eddies induced by the vertical velocity profile of the gas (section
~\ref{sec:dense-midplane-physics}) excite random velocities in the
particles, diffusing the layer and preventing it from settling into a
sufficiently dense state. Detailed two-phase fluid models by
\textit{Cuzzi et al.}  (1993), \textit{Champney et al.} (1995), and
\textit{Dobrovolskis et al.} (1999) confirmed this behavior.

It is sometimes assumed that ongoing, but slow, particle growth to
larger particles, with lower relative velocities and thus thinner
layers (section 3.2), can lead to $\rhop \sim \rho^*$ and
gravitational instability can then occur. However, merely achieving
the formal requirement for marginal gravitational instability does not
inevitably lead to planetesimals.  For particles which are large
enough to settle into suitably dense layers for {\it marginal}
instability under self-generated turbulence (\textit{Weidenschilling},
1980; \textit{Cuzzi et al.}, 1993) random velocities are not damped on
a collapse timescale, so incipient instabilities merely ``bounce'' and
tidally diverge. This is like the behavior seen in Saturn's A ring,
much of which is gravitationally unstable by these same criteria
(\textit{Salo}, 1992; \textit{Karjalainen and Salo}, 2004).  Direct
collapse to planetesimals is much harder to achieve, requiring much
lower relative velocities, and is unlikely to have occurred this way
(\textit{Cuzzi et al.}  1994, \textit{Weidenschilling}, 1995;
\textit{Cuzzi and Weidenschilling}, 2005).  Recent results by
\textit{Tanga et al.} (2004) assume an artificial damping by gas drag
and find gravitationally bound clumps form which, while not collapsing
directly to planetesimals, retain their identity for extended periods,
perhaps allowing for slow shrinkage; this is worth further numerical
modeling with more realistic damping physics, but still presumes a
globally laminar nebula.

For very small particles ($\agr<$ 1mm; the highly relevant chondrule
size), a different type of instability comes into play because the
particles are firmly trapped to the gas by their short stopping times,
and the combined system forms a single ``one-phase'' fluid which is
stabilized against producing turbulence by its vertical density
gradient (\textit{Sekiya}, 1998; \textit{Youdin and Shu}, 2002;
\textit{Youdin and Chiang}, 2004; \textit{Garaud and Lin}, 2004). Even
for midplane layers of such small particles to {\it approach} a
suitable density for this to occur requires nebula turbulence to drop
to what may be implausibly low values ($\alpha < 10^{-8}$ to
$10^{-10}$). Moreover, such one-phase layers, with particle stopping
times $\tstop$ much less than the dynamical collapse time $(G
\rhop)^{-1/2}$, cannot become ``unstable'' and collapse on the
dynamical timescale as normally envisioned, because of {\it gas}
pressure support, which is usually ignored (\textit{Sekiya}, 1983;
\textit{Safronov}, 1991). \textit{Sekiya} (1983) finds that particle
densities must exceed $10^4 \rho^*$ for such particles to undergo
instability and actually collapse. While especially difficult on
one-phase instabilities by definition, this obstacle should be
considered for any particle with stopping time much shorter than the
dynamical collapse time - that is, pretty much anything smaller than a
meter for $\rhop \sim \rho^*$.

A slower ``sedimentation'' from axisymmetric rings (or even localized
blobs of high density, which might form through fragmentation of such
dense, differentially rotating rings), has also been proposed to occur
under conditions normally ascribed to marginal gravitational
instability (\textit{Sekiya}, 1983; \textit{Safronov}, 1991;
\textit{Ward}, 2000), but this effect has only been modeled under
nonturbulent conditions where, as mentioned above, growth can be quite
fast by sticking alone. In a turbulent nebula, diffusion (or other
complications discussed below, such as large vortices, spiral density
waves, etc) might preclude formation of all but the broadest-scale
``rings'' of this sort, which have radial scales comparable to $H$ and
grow only on extremely long timescales.

\subsection{Planetesimal formation in turbulence} 

A case can be made that astronomical, asteroidal, and meteoritic
observations require planetesimal growth to stall at sizes much
smaller than several km, for something like a million years
(\textit{Dullemond and Dominik}, 2005; \textit{Cuzzi and
  Weidenschilling}, 2005; \textit{Cuzzi et al.}, 2005). This is perhaps
most easily explained by the presence of ubiquitous weak turbulence
($\alpha > 10^{-4}$).  Once having grown to meter-size, particles
couple to the largest, most energetic turbulent eddies, leading to
mutual collisions at relative velocities on the order of $
v_\mathrm{turb} \sim \sqrt{\alpha} c \sim 30$ m/s, which are
probably disruptive, stalling incremental growth by sticking at around
a meter in size. Astrophysical observations supporting this inference
are discussed in the next section. In principle, planetesimal
formation could merely await cessation of nebula turbulence and then
happen all at once; pros and cons of this simple concept are discussed
by \textit{Cuzzi and Weidenschilling} (2005).  The main difficulty
with this concept is the very robust nature of growth in dense
midplane layers of nonturbulent nebulae, compared to the very extended
duration of $10^6$ years which apparently characterized meteorite
parent body formation (chapter by \textit{Wadhwa et al.}).
Furthermore, if turbulence merely ceased at the appropriate time for
parent body formation to begin, particles of all sizes would settle
and accrete together, leaving unexplained the very well characterized
chondrite size distributions we observe.  Alternately, several
suggestions have been advanced as to how the meter-sized barrier might
be overcome even in ongoing turbulence, as described below.

\dominiketalsubsubsection{Concentration of boulders in large nebula gas structures} 
The speedy inward radial drift of meter-sized particles in nebulae
where settling is precluded by turbulence might
be slowed if they can be, even temporarily, trapped by one of several
possible fluid dynamical effects. It has been proposed that such
trapping concentrates them and leads to planetesimal growth as well.

Large nebula gas dynamical structures such as systematically rotating
vortices (not true turbulent eddies) have the property of
concentrating large boulders near their centers (\textit{Barge and
  Sommeria}, 1995; \textit{Tanga et al.}, 1996; \textit{Bracco et al.},
1998; \textit{Godon and Livio}, 2000; \textit{Klahr and Bodenheimer},
2006).  In some of these models the vortices are simply prescribed
and/or there is no feedback from the particles.  Moreover, there are
strong vertical velocities present in realistic vortices, and the
vortical flows which concentrate m-size particles are not found near
the midplane, where the m-sized particles reside (\textit{Barranco and
  Marcus}, 2005).  Finally, there may be a tendency of particle
concentrations formed in modeled vortices to drift out of them and/or
destroy the vortex (\textit{Johansen et al.}, 2004).

Another possibility of interest is the buildup of solids near the
peaks of nearly axisymmetric, localized radial pressure maxima, which
might for instance be associated with spiral density waves
(\textit{Haghighipour and Boss}, 2003a,b; \textit{Rice et al.}, 2004).
\textit{Johansen et al.} (2006) noted boulder concentration in radial
high pressure zones of their full simulation, but (in contrast to
above suggestions about vortices), saw no concentration of meter-sized
particles in the closest thing they could resolve in the nature of
actual turbulent eddies.  Perhaps this merely highlights the key
difference between systematically rotating (and often artificially
imposed) vortical fluid structures, and realistic eddies in realistic
turbulence.

Overall, models of boulder concentration in large-scale fluid structures will
need to assess the tendency for rapidly colliding meter-sized particles in such
regions to destroy each other, in the real turbulence which will surely
accompany such structures.  For instance, breaking spiral density waves are
themselves potent drivers for strong turbulence (\textit{Boley et al.}, 2005).

\dominiketalsubsubsection{Concentration of chondrules in 3D turbulence} 
Another suggestion for particle growth beyond a meter in turbulent
nebulae is motivated by observed size-sorting in chondrites.
\textit{Cuzzi et al.} (1996, 2001) have advanced the model of
turbulent concentration of chondrule-sized (mm or smaller diameter)
particles into dense zones, that ultimately become the planetesimals
we observe. This effect, which occurs in genuine, 3D turbulence (both
in numerical models and laboratory experiments), naturally satisfies
meteoritics observations in several ways under quite plausible nebula
conditions.  It offers the potential to leapfrog the problematic
meter-size range entirely and would be applicable (to differing
particle types) throughout the solar system (see \textit{Cuzzi and
  Weidenschilling}, 2005 and \textit{Cuzzi et al.}, 2005 for reviews).
This scenario faces the obstacle that the dense, particle-rich zones
which certainly {\it do} form are far from solid density, and might be
disrupted by gas pressure or turbulence before they can form solid
planetesimals. As with dense midplane layers of small particles, gas
pressure is a formidable barrier to gravitational instability on a
dynamical timescale in dense zones of chondrule-sized particles formed
by turbulent concentration. However, as with other small-particle
scenarios, sedimentation is a possibility on longer timescales than
that of dynamical collapse. It is promising that \textit{Sekiya}
(1983) found that zones of these densities, while ``incompressible''
on the dynamical timescale, form stable modes. Current studies are
assessing whether the dense zones can survive perturbations long
enough to evolve into planetesimals.

\subsection{Summary of the situation regarding planetesimal formation} 

As of the writing of this chapter, the path to planetesimal formation
remains unclear. In nonturbulent nebulae, a variety of options seem to
exist for growth which - while not on dynamical collapse timescales,
is rapid on cosmogonic timescales ($\ll 10^5$ years). However, this
set of conditions and growth timescales seems to be at odds with
asteroidal, meteoritic, and astronomical observations of several kinds
(\textit{Russell et al.}, 2006; \textit{Dullemond and Dominik}, 2005;
\textit{Cuzzi and Weidenschilling}, 2006; \textit{Cuzzi et al.}, 2005;
chapter by \textit{Wadhwa et al.}).  The alternate set of
scenarios - growth beyond a meter or so in size in turbulent nebulae -
are perhaps more consistent with the observations but are still
incompletely developed beyond some promising directions.  The
challenge is to describe quantitatively the rate at which
planetesimals form under these inefficient conditions.

\section{GLOBAL DISK MODELS WITH SETTLING AND AGGREGATION}
\label{sec:global-disk-models}

Globally modeling a protoplanetary disk including dust settling,
aggregation, radial drift and mixing, along with radiative transfer
solutions for the disk temperature and spectrum form a major numerical
challenge, because of the many orders of magnitude that have to be
covered both in time scales (inner disk versus outer disk, growth of
small particles versus growth of large objects) and particle sizes.
Further numerical difficulties result from the fact that small
particles may contribute significantly to the growth of larger bodies,
and careful renormalization schemes are necessary to treat these
processes correctly and in a mass-conserving manner (\textit{Dullemond and
Dominik}, 2005).  Further difficulties arise from uncertainty about the
strength and spatial extent of turbulence during the different
evolutionary phases of a disk.  A complete model covering an entire
disk and the entire growth process along with all relevant disk
physics is currently still out of reach.  Work so far has therefore
either focused on specific locations in the disk, or has used
parametrized descriptions of turbulence with limited sets of physical
growth processes.  However, these ``single slice'' models have the
problem that radial drift can become so large for m-sized objects,
that these leave the slice on a time scale of a few orbital times
(\textit{Weidenschilling}, 1977; section \ref{sec:prior-to-midplane}).
Nevertheless, important results have come forth from these efforts,
that test underlying assumptions of the models.

For the spectral and imaging appearance of disks, there are two main
processes that should produce easily observable results: particle
settling and particle growth.  Particle settling is due to the
vertical component of gravity acting in the disk on the pressure-less
dust component (section \ref{sec:role-of-turb}).  Neglecting growth
for the moment, settling leads to a vertical stratification and size
sorting in the disk.  Small particles settle slowly and should be
present in the disk atmosphere for a long time, while large particles
settle faster and to smaller scale heights.  While in a laminar nebula
this is a purely time dependent phenomenon, this result is permanent
in a turbulent nebula as each particle size is spread over its
equilibrium scale height (\textit{Dubrulle et al.}, 1995).  From a pure
settling model, one would therefore expect that \emph{small dust
  grains} will increasingly dominate dust emission features (cause
strong feature-to-continuum ratios) as large grains disappear from the
surface layers.

Grain growth may have the opposite effect.  While vertical mixing and
settling still should lead to a size stratification, particle growth
can become so efficient that all small particles are removed from the
gas.  In this case, dust emission features should be characteristic
for \emph{larger particles} (i.e. no or weak features, see chapter by
\textit{Natta et al.}).  At the same time, the overall opacity
decreases dramatically.  This effect can become significant, as has
been realized already early on (\textit{Weidenschilling}, 1980, 1984;
\textit{Mizuno}, 1989).  In order to keep the small particle abundance
at realistic levels and the dust opacity high, \textit{Mizuno et al.}
(1988) considered a steady inflow of small particles into a disk.
However, disks with signs of small particles are still observed around
stars that seem to have completely removed their parental clouds, so
this is not a general solution for this problem.  In the following we
discuss the different disk models documented in the literature.  We
begin with a discussion of earlier models focusing on specific regions
of the solar system.

\subsection{Models limited to specific regions in the solar system}

Models considering dust settling and growth in a single vertical slice
have a long tradition, and have been reviewed in previous Protostars
and Planets III (\textit{Cuzzi and Weidenschilling}, 1993).  We therefore
refrain from an in-depth coverage and only recall a few of the main
results.  The global models discussed later are basically similar
calculations, with higher resolution, and for a large set of radii.

\textit{Weidenschilling} has studied the aggregation in laminar
(\textit{Weidenschilling}, 1980, 2000) and turbulent
(\textit{Weidenschilling}, 1984, 1988) nebulae, focusing on the region
of terrestrial planet formation, in particular around 1AU.  These
papers contain the basic descriptions of dust settling and growth
under laminar and turbulent conditions.  They show the occurrence of a
rain-out after particles have grown to sizes where the settling motion
starts to exceed the thermal motions.  \textit{Nakagawa et al.} (1981,
1986) study settling and growth in vertical slices, also concentrating
on the terrestrial planet formation regions.  They find that within
3000 years, the midplane is populated by cm-sized grains.
\textit{Weidenschilling} (1997) studied the formation of comets in the
outer solar system with a detailed model of a non-turbulent nebula,
solving the coagulation equation around 30AU.  In these calculations,
growth initially proceeds by Brownian motion, without significant
settling, for the first 10000yrs.  Then, particles become large enough
and start to settle, so that the concentration of solids increases
quickly after $5\times10^4$yr.  The particle layer reaches the
critical density where the layer gravitational instability is often
assumed to occur, but first the high velocity dispersion prevents the
collapse.  Later, a transient density enhancement still occurs, but
due to the small collisional cross section of the typically 1m-sized
bodies, growth must still happen in individual 2-body collisions.

\subsection{Dust aggregation during early disk evolution}

\textit{Schmitt et al.} (1997) implemented dust coagulation in an
$\alpha$ disk model.  They considered the growth of PCA in a
one-dimensional disk model, i.e. without resolving the vertical
structure of the disk.  The evolution of the dust size distribution is
followed for 100 years only.  In this time, at a radius of 30AU from
the star, first the smallest particles disappear within 10 years, due
to Brownian motion aggregation.  This is followed by a self-similar
growth phase during which the volume of the particles increases by 6
orders of magnitude.  Aggregation is faster in the inner disk, and the
decrease in opacity followed by rapid cooling leads to a \emph{thermal
  gap} in the disk around 3AU.  Using the CCA particles, aggregation
stops in this model after the small grains have been removed.  For
such particles, longer timescales are required to continue the growth.

Global models of dust aggregation during the prestellar collapse stage
and into the early disk formation stage are numerically feasible
because the growth of particles is limited. \textit{Suttner et al.}
(1999) and \textit{Suttner and Yorke} (2001) study the evolution of
dust particles in protostellar envelopes, during collapse, and the
first 10$^4$ years of dynamical disk evolution, respectively.  These
very ambitious models include a radiation hydrodynamic code that can
treat dust aggregation and shattering using an implicit numerical
scheme.  They find that during a collapse phase of 10$^3$ years, dust
particles grow due to Brownian motion and differential radiative
forces, and can be shattered by high velocity collisions cause by
radiative forces.  During early disk evolution, they find that at 30AU
from the star within the first pressure scale height from the
midplane, small particles are heavily depleted because the high
densities lead to frequent collisions.  The largest particles grow by
a factor of 100 in mass.  Similar results are found for PCA particles,
while CCA particles show accelerated aggregation because of the
enhanced cross section in massive particles.  Within 10$^4$ years,
most dust moves to the size grid limit of 0.2mm.  While aggregation is
significant near the midplane (opacities are reduced by more than a
factor 10), the overall structure of the model is not yet affected
strongly, because at the low densities far from the midplane
aggregation is limited and changes in the opacity are only due to
differential advection.

\subsection{Global settling models}

Settling of dust without growth goes much slower than settling that is
accelerated by growth.  However, even pure settling calculations show
significant influence on the spectral energy distributions of disks.
While the vertical optical depth is unaffected by settling alone, the
height at which stellar light is intercepted by the disk surface
changes.  \textit{Miyake and Nakagawa} (1995) computed the effects of dust
settling on the global SED and compared these results with IRAS
observations.  They assume that after the initial settling and growth
phase, enough small particles are left in the disk to provide an
optically thick surface and follow the decrease of the height of this
surface, concluding that this is consistent with the life-times of T
Tauri disks, because the settling time of a 0.1\um{} grain within a
single pressure scale height is of order 10\,Myr.  However, the
initial settling phase does lead to strong effects on the SED, because
settling times at several pressure scale heights are much shorter.

\textit{Dullemond and Dominik} (2004) show that settling from a fully mixed
passive disk leads to a decrease of the surface height in
10$^{4}$--10$^{5}$ years, and can even lead to self-shadowed disks
(see chapter by \textit{Dullemond et al.}).

\begin{figure}[t!]
\epsscale{1.0}
\plotone{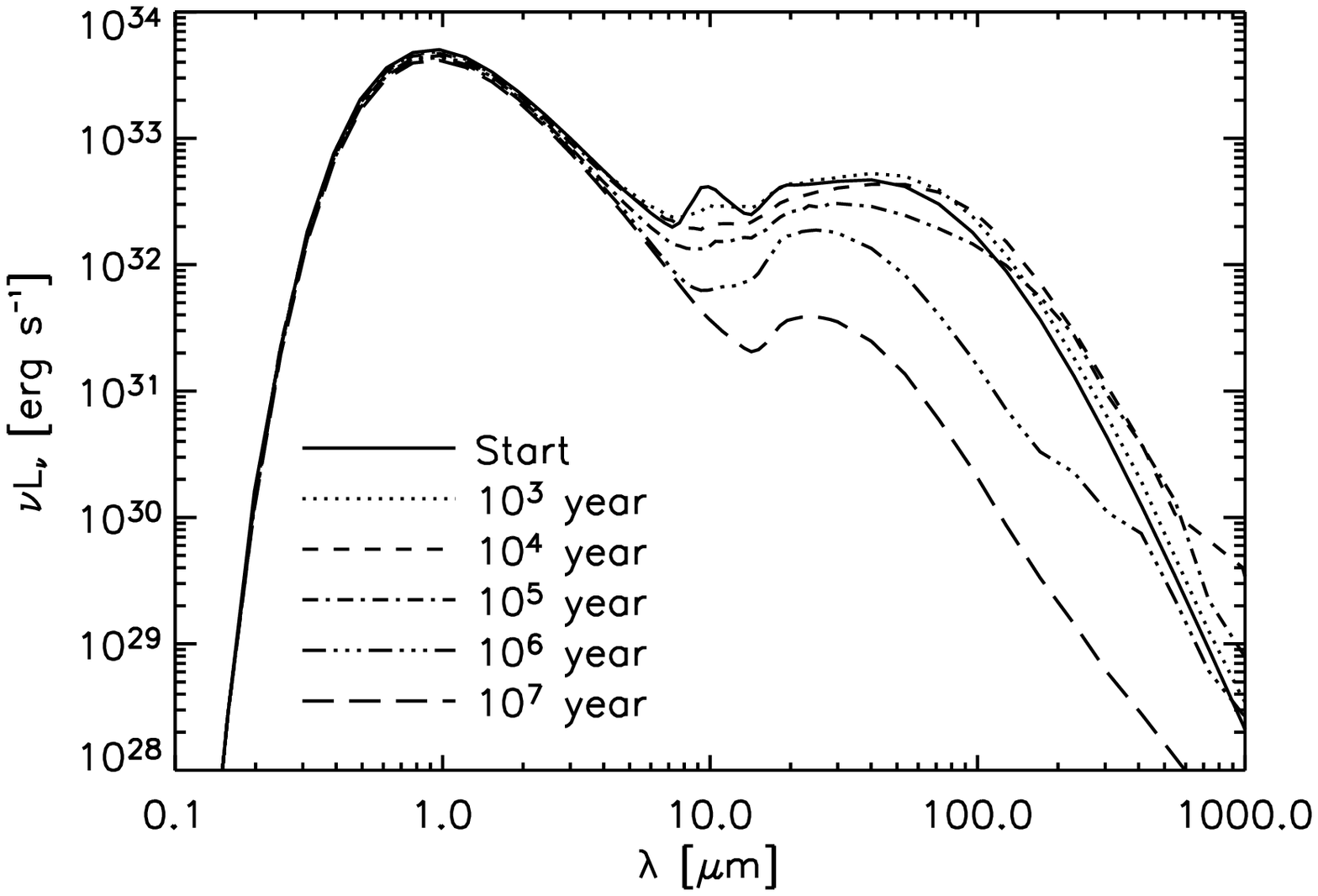}
\plotone{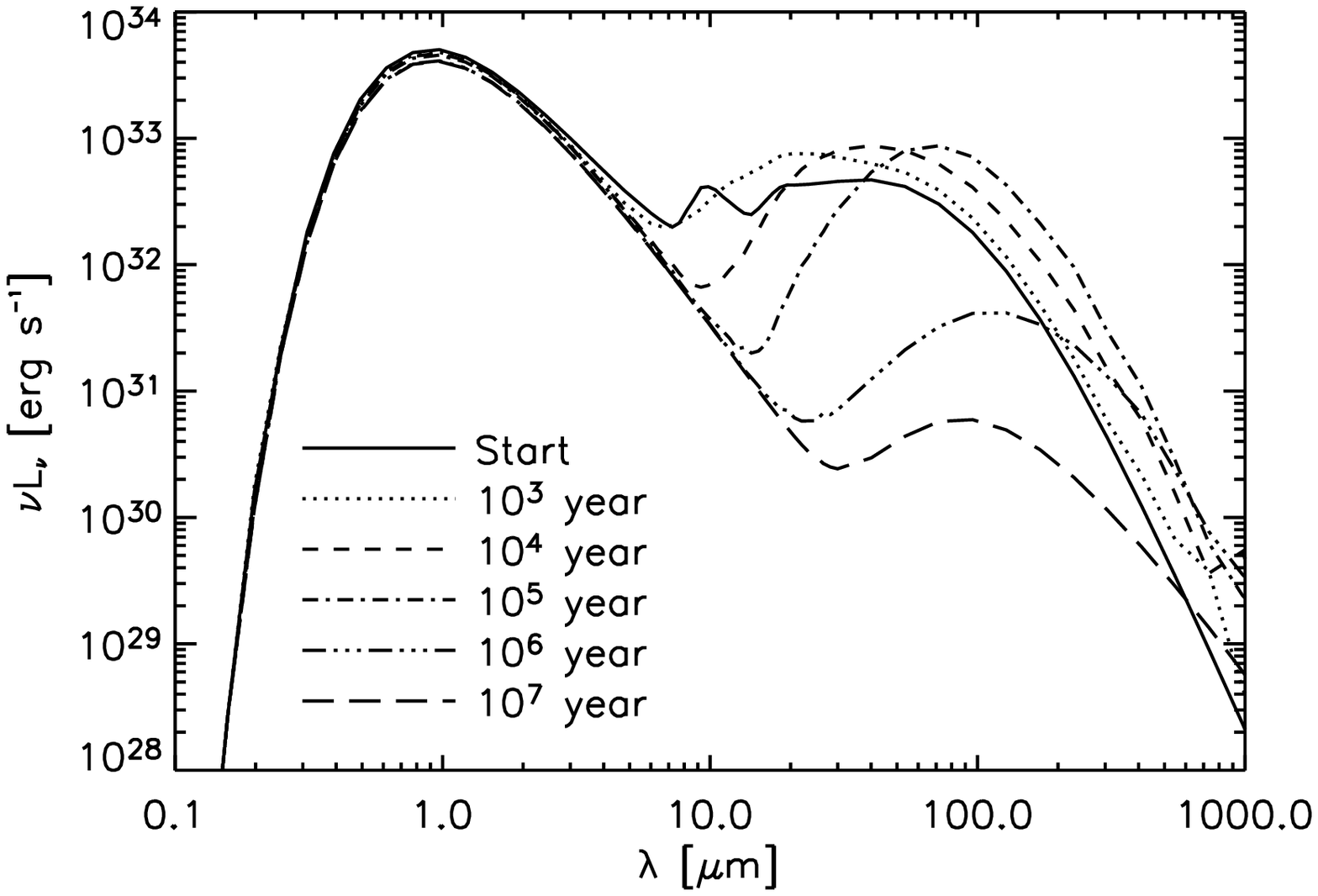}
\caption{\label{fig-sed} \small Time evolution of the disk SED in the
  laminar case (top panel) and the turbulent case (bottom panel).
  From DD05}
\end{figure}

\subsection{Global models of dust growth}
\label{sec:glob-settl-aggr}

\textit{Mizuno} (1989) computes global models including evaporation,
and a steady state assumption using small grains continuously raining
down from the ISM.  The vertical disk structure is not resolved, only
a single zone in the midplane is considered.  He finds that the
Rosseland mean opacity decreases, but then stays steady due to the
second generation grains.

\textit{Kornet et al.} (2001) model the global gas and dust disk by
assuming that at a given radius, the size distribution of dust
particles (or planetesimals) is exactly monodisperse, avoiding the
numerical complications of a full solution of the Smoluchowski
equation.  They find that the distribution of solids in the disk after
10$^7$ years depends strongly on the initial mass and angular momentum
of the disk.

\textit{Ciesla and Cuzzi} (2005) model the global disk using a
four-component model: Dust grains, m-size boulders, planetesimals and
the disk gas.  This model tries to capture the main processes
happening in a disk: growth of dust grains to m-sized bodies, the
migration of m-sized bodies and the resulting creation of evaporation
fronts, and the mixing of small particles and gas by turbulence.  The
paper focuses on the distribution of water in the disk, and the dust
growth processes are handled by assuming timescales for the conversion
from one size to the next.  Such models are therefore mainly useful
for the chemical evolution of the nebula and need detailed aggregation
calculations as input.

The most complete long-term integrations of the equations for dust
settling and growth are described in recent papers by \textit{Tanaka
  et al.}  (2005, henceforth THI05) and \textit{Dullemond and Dominik}
(2005, henceforth DD05).  These papers implement dust settling and
aggregation in individual vertical slices through a disk, and then use
many slices to stitch together an entire disk model, with predictions
for the resulting optical depth and SED from the developing disk.
Both models have different limitations. THI05 consider only laminar
disk models, so that turbulent mixing and collisions between particles
driven by turbulence are not considered. Their calculations are
limited to compact solid particles.  DD05's model is incomplete in
that it does not consider the contributions of radial drift and
differential angular velocities between different particles.  But in
addition to calculations for a laminar nebula, they also introduce
turbulent mixing and turbulent coagulation, as well as PCA and CCA
properties for the resulting dust particles.  THI05 use a two-layer
approximation for the radiative transfer solution, while DD05 run a 3D
Monte-Carlo radiative transfer code to compute the emerging spectrum
of the disk.  Both models find that aggregation proceeds more rapidly
in the inner regions of the disk than in the outer regions, quickly
leading to a region of low optical depth in the inner disk.

Both calculations find that a bi-modal size distribution is formed,
with large particles in the midplane, formed by rainout (the fast
settling of particles after their settling time has decreased below
their growth time) and continuing to grow quickly, and smaller
particles remaining higher up in the disk and then slowly trickling
down.  In the laminar disk, growth stops in the DD05 calculations at
cm sizes because radial drift was ignored.  in THI05, particles
continue to grow beyond this regime.

The settling of dust  causes the surface height of the disk to
decrease, reducing the overall capacity of the disk to reprocess
stellar radiation.  THI05 find that at 8 AU from the star the optical
depth of the disk at 10\um{} reaches about unity after a bit less than
10$^6$yrs.  In the inner disk, the surface height decreases to almost
zero in less than 10$^6$yrs.  The SED of the model shows first a
strong decrease at wavelength of 100\um{} and longer, within the first
10$^5$yrs.  After that the near-IR and mid-IR radiation also decreases
sharply.  THI05 consider their results to be roughly consistent with the
observations of decreasing fluxes at near-IR and mm wavelengths in
disks.

The calculations by DD05 show a more dramatic effect, as shown in
Fig.\ref{fig-sed}.  In the calculations for a laminar disk, here the
surface height already significantly decreases in the first 10$^4$
years, then the effect on the SED is initially strongest in the mid-IR
region.  After 10$^6$yrs, the fluxes have dropped globally by at least
a factor of 10, except for the mm regime, which is affected greatly
only after a few times 10$^6$ years.

In the calculations for a turbulent disk (DD05), the depletion of
small grains in the inner disk is strongly enhanced.  This result is
caused by several effects.  First, turbulent mixing keeps the
particles moving even after they have settled to the mid-plane,
allowing them to be mixed up and rain down again through a cloud of
particles.  Furthermore, vertical mixing in the higher disk regions
mixes low-density material down to higher densities, where aggregation
can proceed much faster.  The material being mixed back up above the
disk is then largely deprived of solids, because the large dust
particles decouple from the gas and stay behind, settling down to the
mid plane.  The changes to the SED caused by coagulation and settling
in a turbulent disk are dramatic, and clearly inconsistent with the
observations of disks around T Tauri stars that indicate lifetimes of
up to 10$^7$ years.  DD05 conclude that ongoing particle destruction
must play an important role, leading to a steady-state size
distribution for small particles (section \ref{physical properties of
  aggregates}).

\subsection{The role of aggregate structure}

Up to now, most solutions for the aggregation equation in disks are
still based on the assumption of compact particles resulting from the
growth process.  However, at least for the small aggregates formed
initially, this assumption is certainly false.  First of all, if
aggregates are fluffy, with large surface-to-mass ratios, it will be
much easier to keep these particles in the disk surface where they can
be observed as scattering and IR emitting grains.  Observations of the
10\um{} silicate features show that in many disks, the population
emitting in this wavelength range is dominated by particles larger
than interstellar (\textit{van Boekel et al.}, 2005;
\textit{Kessler-Silacci et al.}, 2005).  When modelled with compact
grains, the typical size of such grains is several microns, with
corresponding settling times less than a Myr.  When modelled with
aggregates, particles have to be much larger to produce similar
signatures (flattened feature shapes, e.g. \textit{Min et al.}, 2005).

When considering the growth time scales, in particular in regions
where settling is driving the relative velocities, the timescales are
surprisingly similar to the case of compact particles
(\textit{Safronov}, 1969, \textit{Weidenschilling}, 1980).  While
initially, fluffy particles settle and grow slowly because of small
settling velocities, the larger collisional cross section soon leads to
fast collection of small particles, and fluffy particles reach the
mid-plane as fast as compact grains, and with similar masses
collected.

\def\twidth{5.3cm}
\def\titema{$-$ }
\def\titem{\newline$-$ }
\begin{table*}[t!]
\caption{Overview}
\label{tab:overview}
\small
\begin{tabular}[t]{|p{\twidth}|p{\twidth}|p{\twidth}|}
\hline
\multicolumn{1}{|c}{\textbf{What We Really Know}}&
\multicolumn{1}{|c}{\textbf{Main Controversies/Questions}}&
\multicolumn{1}{|c|}{\textbf{Future Priorities}}\\\hline
\multicolumn{3}{|c|}{Microphysics}\\\hline
\titema Dust particles stick in collisions with
  less than $\sim 1$ m/s velocity due to van der Waals force or hydrogen
  bonding.
\titem For low relative velocities ($\ll 1$m/s) a cloud of dust particles
  evolves into fractal aggregates ($\Df<2$) with a quasi-monodisperse mass
  distribution.
\titem Due to the increasing collision energy, growing fractal
  aggregates can no longer keep their structures so that non-fractal
  (but very porous) aggregates form (still at $v\ll 1$m/s).
\titem Macroscopic aggregates have porosities $>65$\% when
  collisional compaction, and not sintering or melting occurs.
&
\titema At what aggregate size does compaction happen in a nebula
environment?
\titem When do collisions between macroscopic aggregates result in
  sticking?  Some experiments show no sticking at rather low impact
  velocities, while others show sticking at high impact speeds.
\titem How important are special material properties: organics,
  ices, magnetic and electrically charged particles?
\titem What are the main physical parameters (e.g. velocity, impact
angle, aggregate porosity/material/shape/mass) determining the 
outcome of a collision?
 &
\titema More empirical studies in collisions between macroscopic
  aggregates required.
\titem Macroscopic model for aggregate collisions (continuum description)
  based on microscopic model and experimental results.
\titem Develop recipes for using the microphysics in large scale
aggregation calculations.
\titem Develop aggregation models that treat aggregate structure as
a \emph{variable} in a self-consistent way.
\\\hline
\multicolumn{3}{|c|}{Nebula processes}\\\hline
\titema Particle velocities and relative velocities in turbulent and
nonturbulent nebulae are understood; values are $<1$ m/s for $\agr\rho
< 1-3$ g cm$^{-2}$ depending on alpha.
\titem Radial drift decouples large amounts of solids from the gas and
migrates it radially, changing nebula mass distribution and chemistry
\titem Turbulent diffusion can offset inward drift for particles of cm size
and smaller, relieving the ``problem'' about age differences between
CAIs and chondrules.
&
\titema What happens to dust aggregates in highly mass-loaded
  regions in the solar nebula, e.g. midplane, eddies, stagnation
  points?
\titem Is the nebula turbulent? If so, how does the intensity vary with
  location and time?  Can purely hydrodynamical processes produce
  self-sustaining turbulence in the terrestrial planet formation zone?
\titem Can large-scale structures (vortices, spiral density waves)
  remain stable long enough to concentrate boulder-size particles?
\titem Can dense turbulently concentrated zones of chondrule-size particles
 survive to become actual planetesimals?
&
\titema Relative velocities in highly mass-loaded regions in the solar
  nebula, e.g. midplane, eddies, stagnation points.
\titem Improve our understanding of turbulence production processes at very
  high nebula Reynolds numbers.
\titem Model effects of MRI-active upper layers on dense, non-ionized gas in
magnetically dead zones.
\titem Model the evolution of dense strengthless clumps of particles in turbulent gas.
\titem Model collisional processes in boulder-rich vortices and high-pressure
zones.
\titem Model evolution of dense clumps in turbulent gas.
\\\hline
\multicolumn{3}{|c|}{Global modelling and comparison with observations}\\\hline
\titema Small grains are quickly depleted by
incorporation into larger grains.
\titem Growth timescales are short for small compact and fractal grains alike.
\titem Vertical mixing and small grain replenishment are necessary to keep the
observed disk structures (thick/flaring).
&
\titema What is the role of fragmentation for the small grain
component?
\titem Are the ``small'' grains seen really large, fluffy aggregates?
\titem Are the mm/cm sized grains seen in observations compact
  particles, or much larger fractal aggregates?
\titem What is the global role of radial transport?
&
\titema Study the optical properties of \emph{large} aggregates, fluffy and
  compact.
\titem Implement realistic opacities in disk models to produce
  predictions and compare with observations.
\titem Construct truly global models including radial transport.
\titem More resolved disk images at many wavelengths, to better
  constrain models.
\\\hline
\end{tabular}
\end{table*}

\section{SUMMARY AND FUTURE PROSPECTS}

A lot has been achieved in the last few years, and our understanding
of dust growth has advanced significantly.  There are a number of
issues where we now have clear answers.  However, a number of major
controversies remain, and future work will be needed to address these
before we can come to a global picture of how dust growth in
protoplanetary disks proceeds and which of the possible ways toward
planetesimals are actually used by nature.  In
table~\ref{tab:overview} on the following page we summarize our main
conclusions and questions, and note some priorities for research in the
near future.

\textbf{ Acknowledgments.} We thank the referee (Stu Weidenschilling)
for valuable comments on the manuscript.  This work was partially
supported by JNC's grant from the Planetary Geology and Geophysics
program.  CD thanks Kees Dullemond for many discussions, Dominik
Paszun for preparing figure 1.  JNC thanks Andrew Youdin for a useful
conversation regarding slowly evolving, large scale structures.

\bigskip
\centerline\textbf{REFERENCES}
\bigskip
\parskip=0pt
{ \small
 \baselineskip=11pt
\def\lpsc{Lunar Planet. Sci.}
\refs Adachi I., Hayashi C., and Nakazawa K. (1976) 
{\em Prog. Theor. Phys., 56}, 1756-1771.

\refs A'Hearn M. A. and the Deep Impact Team  (2005) Deep Impact: The 
Experiment; 37th DPS, Cambridge, England, paper \#35.02.

\refs Barge P., and Sommeria J. (1995) {\em \aap, 296},
L1-L4.

\refs Barranco J. A. and  Marcus P. S. (2005) 
{\em \apj, 623}, 1157-1170.

\refs Bel N. and Schatzman E. (1958) {\em Rev. Mod. Phys., 30}, 1015-1016.

\refs Blum J. (2004) in: {\it Astrophysics of Dust}, 
(A. Witt, G. Clayton and B. Draine, eds.), ASP Conference Series, Vol. 309,
pp. 369-391.

\refs Blum J. (2006) Advances in Physics, submitted. 

\refs Blum J. and M\"unch M. (1993) {\em \icarus, 106}, 151-167.


\refs Blum J. and Schr\"apler R. (2004) {\em \prl, 93}, 115503.

\refs Blum J. and Wurm G. (2000) {\em \icarus, 143}, 138-146.

\refs Blum J., Wurm G., Kempf S., Poppe T., Klahr H., {\it et al.}
(2000) {\em \prl, 85}, 2426-2429.

\refs Blum J., Wurm G., Poppe T., and Heim L.-O. (1999) {\em Earth,
Moon, and Planets, 80}, 285.

\refs Bockel\'ee-Morvan D., Gautier D., Hersant F.,
Hur\'e; J.-M., and Robert F. (2002), {\em \aap, 384}, 1107-1118.

\refs van Boekel R., Min M., Waters L. B. F. M., de Koter A., Dominik 
C., van den Ancker M. E., and Bouwman J. (2005), {\em \aap, 437}, 189-208 .

\refs Boley A. C., Durisen R. H., and Pickett M. K. (2005), {\em ASP 
Conf.~Ser.~341: Chondrites and the Protoplanetary Disk,  341}, 839-848.

\refs Bracco A., Provenzale A., Spiegel E. A., and Yecko P. (1998) 
in {\em Theory of Black Hole Accretion Disks} (M. A. Abramowicz,
G. Bjornsson, and J. E. Pringle, eds.), pp. 254 
Cambridge University Press.

\refs Bridges F. G., Supulver K. D., and  Lin D. N. C. (1996)
{\em \icarus, 123}, 422-435.

\refs Champney J. M., Dobrovolskis A. R., and Cuzzi J. N. (1995) {\em 
Physics of Fluids, 7}, 1703-1711.

\refs Chandrasekhar S. (1961) Hydrodynamic and Hydromagnetic
Stability, p. 589, Oxford University Press.

\refs Chokshi A., Tielens A. G. G. M., and Hollenbach D. (1993)
{\em \apj, 407}, 806-819.

\refs Ciesla F. J. and Cuzzi J. N. (2005) 
 \icarus, in press; also 36th LPSC CDROM. 
 
\refs Colwell J. E. (2003) {\em \icarus, 164}, 188-196.


\refs Cuzzi J. N. (2004) 
{\em \icarus, 168}, 484-497.
 
\refs Cuzzi J. N. and Hogan R. C. (2003) 
{\em \icarus, 164}, 127-138.

\refs Cuzzi J. N. and Weidenschilling S. J. (2006)
in {\em Meteorites and the Early Solar System II} (D. Lauretta, L. A.
Leshin, and H. McSween, eds), Univ of Arizona Press; Tuscon, in press.

\refs Cuzzi J. N. and Zahnle K. (2004) {\em \apj, 614}, 490-496.

\refs Cuzzi J. N., Ciesla F. J., Petaev M. I., Krot A. N., Scott E. R. 
D., and Weidenschilling S. J. (2005) {\em ASP Conf.~Ser.~341: Chondrites 
and the Protoplanetary Disk,  341}, 732-773.

\refs Cuzzi J. N., Davis S. S., and Dobrovolskis A. R. (2003) 
{\em \icarus, 166}, 385-402.

\refs Cuzzi J. N., Dobrovolskis A.R., and Champney J. M. (1993) 
{\em \icarus, 106}, 102-134.

\refs Cuzzi J. N., Dobrovolskis A. R., and Hogan R. C. (1994) 
{\em \lpsc, XXV}, 307-308, LPI, Houston.

\refs Cuzzi J. N., Dobrovolskis A. R., and Hogan R. C. (1996)
in {\em Chondrules and the Protoplanetary Disk} (R.  Hewins, R. Jones,
and E. R. D. Scott, eds), pp. 35-44, Cambridge Univ. Press.

\refs Cuzzi J. N., Hogan R. C., Paque J. M., and Dobrovolskis A. R. (2001)
{\em \apj, 546}, 496-508.

\refs Cyr K., Sharp C. M., and Lunine J. I. (1999)
{\em \jgr, 104}, 19003-19014.
 
\refs Dahneke B. E. (1975) {\em J. Colloid Interf. Sci., 1}, 58-65.
 
\refs Davidsson B. J. R. (2006) {\em Advances in Geosciences}, in press.

\refs Derjaguin B. V., Muller V. M., and Toporov Y. P. (1975)
{\em J. Colloid Interf. Sci., 53}, 314-326.
 
\refs Desch S. J. and Cuzzi J. N. (2000) {\em Icarus,  143}, 87-105.

\refs Dobrovolskis A. R., Dacles-Mariani J. M., and Cuzzi J. N. (1999)
{\em J.G.R. Planets, 104}, E12, 30805-30815.

\refs Dominik C. and N\"ubold H. (2002) {\em Icarus,  157}, 173-186.

\refs Dominik C. and Tielens A. G. G. M. (1995) {\em Phil. Mag.,
  72}, 783-803.

\refs Dominik C. and Tielens A. G. G. M. (1996) {\em Phil. Mag.,
  73}, 1279-1302.

\refs Dominik C. and Tielens A. G. G. M. (1997) {\em \apj, 480}, 647-673.

\refs Dubrulle B., Morfill G. E., and Sterzik M. (1995) 
{\em \icarus, 114}, 237-246.

\refs Dullemond C. P. and Dominik C. (2004), {\em \aap, 421}, 1075-1086.

\refs Dullemond C. P. and Dominik C. (2005), {\em \aap, 434}, 971-986.

\refs Gail H.-P. (2004) {\em \aap, 413}, 571-591.

\refs Garaud P. and Lin D. N. C. (2004) {\em \apj,  608}, 1050-1075.

\refs Godon P. and Livio M. (2000) 
{\em \apj, 537}, 396-404.

\refs Goldreich P. and Lynden-Bell D. (1965) 
{\em \mnras, 130}, 97-124.

\refs Goldreich P. and Ward W. R. (1973) 
{\em \apj, 183}, 1051-1061.

\refs Goodman J. and Pindor B. (2000) 
{\em \icarus, 148}, 537-549.

\refs Greenberg J. M., Mizutani H., and Yamamoto T. (1995) 
{\em \aap, 295}, L35-L38.

\refs Gurevich L. E. and Lebedinsky A. I. (1950) {\em Izvestia Academii
Nauk USSR, 14}, 765. 

\refs Gustafson B.~{\AA}.~S. and Kolokolova L. (1999) {\em \jgr,
  104}, 31711--31720.

\refs Haghighipour N. and Boss A. P. (2003a) 
{\em \apj, 583}, 996-1003. 

\refs Haghighipour N. and Boss A. P. (2003b) 
{\em \apj, 598}, 1301-1311.



\refs Heim L., Blum J.,  Preuss M., and Butt H.-J. (1999) {\em
  \prl, 83}, 3328-3331.

\refs Heim L., Butt H.-J.,  Schr\"apler R., and Blum J. (2005) {\em
Aus. J. Chem., 58}, 671-673. 

\refs Henning T. and Stognienko R. (1996) {\em \aap, 311}, 291-03

\refs Ivlev A. V., Morfill G. E., and Konopka U. (2002) {\em \prl,
  89}, 195502 

\refs Jeans J. H. (1928) Astronomy and Cosmogony, Cambridge University 
Press, p. 337.

\refs Johansen A. and Klahr H. (2005) {\em \apj, 634}, 1353-1371. 

\refs Johansen A., Andersen A. C., and Brandenburg A. (2004)
{\em \aap, 417}, 361-374  .

\refs Johansen A., Klahr H., and Henning Th. (2006) 
{\em \apj, 636}, 1121-1134.

\refs Karjalainen R. and Salo H. (2004) {\em \icarus, 172}, 328-348. 

\refs Kempf S., Pfalzner S., and Henning Th. (1999) {\em \icarus, 141}, 388

\refs Kessler-Silacci J. E., Hillenbrand L. A., Blake G. A., and Meyer 
M. R. (2005), {\em \apj, 622}, 404-429.


\refs Klahr H. and Bodenheimer P. (2006) {\em \apj, 639}, 432-440.

\refs Konopka U., Mokler F., Ivlev A. V., Kretschmer M., Morfill,
G.E., et al.
(2005) {\em New Journal of Physics, 7, 227}, 1-11.

\refs Kornet K., Stepinski T. F., and R\'ozyczka M.
(2001) {\em \aap, 378}, 180-191.
 
\refs Kouchi A., Kudo T., Nakano H., Arakawa M., Watanabe N.,
 Sirono S.-I., Higa M., and Maeno N. (2002) {\em \apj, 566}, L121-L124.

\refs Kozasa T., Blum J., and Mukai T. (1992) {\em \aap, 263}, 423-432.

\refs Krause M. and Blum J. (2004) {\em \prl, 93}, 021103.

\refs Krot A. N., Hutcheon I. D., Yurimoto H., Cuzzi J. N.,
McKeegan K. D., Scott E. R. D., Libourel G., Chaussidon M., Aleon
J., and Petaev M. I. (2005)
{\em \apj,},  622, 1333-1342. 

\refs K\"unzli S. and Benz W. (2003) {\em Meteoritics \& Planetary 
Science, 38}, 5083.
 
\refs Love S. G. and Pettit D. R. (2004) {\em \lpsc, XXXV}, Abstract
\#1119, LPI, Houston (CD-ROM).
 
\refs Markiewicz W. J., Mizuno H., and V{\"o}lk H. J.  (1991)
{\em \aap, 242}, 286-289.

\refs Marshall J. and Cuzzi J. (2001) {\em \lpsc, XXXII}, 1262, LPI,
Houston.

\refs Marshall J., Sauke T. A., and Cuzzi J. N. (2005) {\em \grl, 32},
L11202-L11205.

\refs McCabe C., Duch{\^e}ne G., and Ghez A. M. (2003) {\em \apj,
  588}, L113.

\refs Miyake K. and Nakagawa Y. (1995) {\em \apj,  441}, 361-384.
 
\refs Mizuno H. (1989) {\em Icarus, 80}, 189-201.
 
\refs Mizuno H., Markiewicz W. J., and Voelk H. J. (1988) {\em \aap,
195}, 183-192.
 
\refs Morfill G. E. and V{\"o}lk H. J. (1984)  
{\em \apj, 287}, 371-395.

\refs Nakagawa Y., Nakazawa K., and Hayashi C. (1981) {\em \icarus, 
  45}, 517-528.
 
\refs Nakagawa Y., Sekiya M., and Hayashi C. (1986) 
{\em \icarus, 67}, 375-390.

\refs N\"ubold H. and Glassmeier K.-H. (2000) {\em Icarus,  144}, 149-159.

\refs N\"ubold H., Poppe T., Rost M., Dominik C., and Glassmeier K.-H. 
(2003) {\em Icarus,  165}, 195-214.

\refs Nuth J. A. and Wilkinson G. M. (1995) {\em Icarus,  117}, 431-434.
 
\refs Nuth J. A., Faris J., Wasilewski P., and Berg O. (1994) {\em 
Icarus,  107}, 155-163.

\refs Ossenkopf V. (1993) {\em \aap, 280}, 617-646.

\refs Paszun D. and Dominik C. (2006) \icarus, in press.


\refs Poppe T. and Schr\"apler R. (2005) {\em \aap, 438}, 1-9.

\refs Poppe T., Blum J., and Henning Th. (2000a) {\em \apj, 533}, 454-471.

\refs Poppe T., Blum J., and Henning Th. (2000b) {\em \apj, 533}, 472-480.

\refs Rice W. K. M., Lodato G.,  Pringle J. E.,  Armitage P. J., and
Bonnell I. A. (2004) {\em \mnras, 355}, 543-552.

\refs Russell S. S., Hartmann L. A., Cuzzi J. N., Krot A. N., and
Weidenschilling S. J. (2006) 
in {\em Meteorites and the Early Solar System, II} (D. Lauretta, L. A.
Leshin, and H.  McSween, eds.), in press.

\refs Safronov V. S. (1960) {\em Annales d'Astrophysique, 23}, 979-982.

\refs Safronov V. S. (1969)
NASA TTF-677.

\refs Safronov V. S. (1991)
{\em \icarus, 94}, 260-271.

\refs Salo H. (1992) {\em \nat, 359}, 619-621.

\refs Schmitt W., Henning Th., and Mucha R. (1997) {\em \aap, 325}, 569-584.
 
\refs Scott E. R. D., Love S. G., and Krot A. N. (1996) 
in {\em Chondrules and the Protoplanetary Disk} (R. Hewins, R. Jones, and
E. R. D. Scott, eds), pp. 87-96, Cambridge Univ. Press.

\refs Sekiya M. (1983)
{\em Prog. Theor. Physics, 69}, 1116-1130.

\refs Sekiya M. (1998) 
{\em \icarus, 133}, 298-309.

\refs Sekiya M. and Takeda H. (2003) {\em Earth, Planets, and Space,
55}, 263-269.

\refs Sekiya M. and Takeda H. (2005) {\em \icarus, 176}, 220-223

\refs Shakura N. I., Sunyaev R. A., and Zilitinkevich S. S. (1978)
{\em \aap, 62}, 179-187.

\refs Sirono S. (2004) {\em \icarus, 167}, 431-452.

\refs Sirono S. and Greenberg J. M. (2000) {\em \icarus, 145}, 230-238.

\refs v. Smoluchowski M. (1916) {\em Physik. Zeit., 17}, 557-585.

\refs Stepinski T. F. and Valageas P. (1996) 
{\em \aap, 309}, 301-312.

\refs Stepinski T. F. and Valageas P. (1997) 
{\em \aap, 319}, 1007-1019.

\refs Stone J. M., Gammie C. F., Balbus S. A., and Hawley J. F.
(2000)
in {\em Protostars and Planets IV} (V. Mannings, A. P. Boss, and
S. S. Russell, eds.), pp. 589-599;  Univ. of Arizona Press, Tuscon.

\refs Supulver K. and Lin D. N. C. (2000) 
        {\em \icarus, 146}, 525-540.

\refs Suttner G. and Yorke H. W. (2001) {\em \apj, 551}, 461-477.
 
\refs Suttner G., Yorke H. W., and Lin D. N. C. (1999), {\em \apj, 
524}, 857-866.

\refs Tanaka H., Himeno Y., and Ida S. (2005), {\em \apj, 625}, 414-426.

\refs Tanga P., Babiano A., Dubrulle B., and Provenzale A. (1996) 
{\em \icarus, 121}, 158-170.

\refs Tanga P., Weidenschilling S. J., Michel P., and 
Richardson D. C. (2004) {\em \aap, 427}, 1105-1115.

\refs Toomre A. (1964)
{\em \apj, 139}, 1217-1238.

\refs T\'unyi I., Guba P., Roth L. E., and Timko M. (2003) {\em
  Earth Moon and Planets, 93}, 65-74.

\refs Ueta T. and Meixner M. (2003) {\em \apj, 586}, 1338-1355.

\refs V\"{o}lk H. J., Jones F. C., Morfill G. E., and R{\"o}ser
S. (1980) 
{\em \aap, 85}, 316-325.

\refs Valverde J. M., Quintanilla M. A. S., and Castellanos A.
(2004) {\em \prl, 92}, 258303

\refs Ward W. R. (1976) in {\em Frontiers of Astrophysics}
(E. H. Avrett, ed.), pp. 1-40.

\refs Ward W. R. (2000) 
in {\em Origin of the Earth and Moon} (R. M. Canup et al., eds.),
pp. 75-84, Univ. Arizona Press, Tucson.

\refs Wasson J. T., Trigo-Rodriguez J. M., and Rubin A. E. (2005)
{\em \lpsc, XXXVI}, Abstract \#2314, LPI, Houston.

\refs Watson P. K., Mizes H., Castellanos A., and P\'erez A.
(1997) in: {\it Powders \& Grains 97}, (R. Behringer and
J. T. Jenkins, eds), pp. 109-112, Balkema, Rotterdam.

\refs Weidenschilling S. J. (1977) 
{\em \mnras, 180}, 57-70.

\refs Weidenschilling S. J. (1980)
{\em \icarus, 44}, 172-189.   

\refs Weidenschilling S. J. (1984), {\em \icarus, 60}, 553-567.

\refs Weidenschilling S. J. (1988) in {\em Meteorites and the early solar system} 
(J. A. Kerridge and M. S. Matthews, eds), pp. 348-371, University of Arizona
Press, Tuscon.

\refs Weidenschilling S. J. (1995)
{\em \icarus, 116}, 433-435.

\refs Weidenschilling S. J. (1997) 
{\em \icarus, 127}, 290-306. 

\refs Weidenschilling S. J. (2000) 
{\em \ssr, 92}, 295-310.

\refs Weidenschilling S. J. and  Cuzzi J. N. (1993) Protostars and
Planets III, 1031-1060.

\refs Whipple F. (1972) 
in {\em From Plasma to Planet} (A. Elvius, ed.), pp. 211-232, Wiley, New York.

\refs Wolf S. (2003) {\em \apj, 582}, 859-868.


\refs Wurm G. and Blum J. (1998) {\em \icarus, 132}, 125-136.

\refs Wurm G. and Schnaiter M. (2002) {\em \apj, 567}, 370-375.

\refs Wurm G., Blum J., and Colwell J. E. (2001) {\em \icarus,
  151}, 318-321. 
\refs Wurm G., Paraskov G., and Krauss O. (2004b) {\em \apj,
  606}, 983-987. 

\refs Wurm G., Paraskov G., and Krauss O. (2005) {\em \icarus, 178}, 
253-263.

\refs Wurm G., Relke H., Dorschner J., and Krauss O. (2004a)
{\em J. Quant. Spect. Rad. Transf., 89}, 371-384.

\refs Youdin A. N. (2004) {\em ASP Conf. Ser., 323}, 319-327.

\refs Youdin A. N. and Chiang E. I. (2004) {\em \apj, 601}, 1109-1119.
 
\refs Youdin A. and Shu F. (2002) 
{\em \apj, 580}, 494-505.

\refs Yurimoto H. and Kuramoto N. (2004)
{\em Science, 305}, 1763-1766.

} 

\end{document}